\input amstex
\input psfig.sty
\magnification 1200
\TagsOnRight
\def\qed{\ifhmode\unskip\nobreak\fi\ifmmode\ifinner\else
 \hskip5pt\fi\fi\hbox{\hskip5pt\vrule width4pt
 height6pt depth1.5pt\hskip1pt}}
 \def\adots{\mathinner{\mkern2mu\raise1pt\hbox{.}
\mkern3mu\raise4pt\hbox{.}\mkern1mu\raise7pt\hbox{.}}}
\def\sdots{\mathinner{
     \mskip.01mu\raise1pt\vbox{\kern1pt\hbox{.}}
     \mskip.01mu\raise3pt\hbox{.}
     \mskip.01mu\raise5pt\hbox{.}
     \mskip1mu}}
\NoBlackBoxes
\baselineskip 20 pt
\parskip 5 pt

\centerline {\bf EXACT SOLUTIONS TO THE FOCUSING}
\centerline {\bf NONLINEAR SCHR\"ODINGER EQUATION}

\vskip 10 pt
\centerline {Tuncay Aktosun}
\vskip -6 pt
\centerline {Department of Mathematics}
\vskip -6 pt
\centerline {University of Texas at Arlington}
\vskip -6 pt
\centerline {Arlington, TX 76019-0408, USA}

\centerline{Francesco Demontis and Cornelis van der Mee}
\vskip -6 pt
\centerline{Dipartimento di Matematica e Informatica}
\vskip -6 pt
\centerline{Universit\`a di Cagliari}
\vskip -6 pt
\centerline{Viale Merello 92, 09123 Cagliari, Italy}

\vskip 10 pt

\noindent {\bf Abstract}:
A method is given to construct globally analytic (in space and time)
exact solutions to the focusing cubic
nonlinear Schr\"odinger equation on the line.
An explicit formula and its equivalents are presented to
express such exact solutions in a compact form in terms of
matrix exponentials. Such exact solutions can alternatively be
written explicitly as
algebraic combinations of exponential, trigonometric, and polynomial
functions of the spatial and temporal coordinates.


\vskip -6 pt
\par\noindent {\bf Short title:} Exact solutions to the NLS equation
\newpage

\noindent {\bf 1. INTRODUCTION}
\vskip 3 pt

Consider the focusing cubic nonlinear Schr\"odinger (NLS) equation
$$iu_t+u_{xx}+2|u|^2u=0,\tag 1.1$$
where the subscripts denote the appropriate partial derivatives. The
NLS equation is important for many reasons [1-3,5,31,37]. It arises
in many application areas such as wave propagation in nonlinear
media [37], surface waves on sufficiently deep waters [36], and
signal propagation in optical fibers [24-26]. It was also the second
nonlinear partial differential equation (PDE) whose initial value
problem was discovered [37] to be solvable via the inverse
scattering transform (IST) method.

In this paper we present a method to construct certain exact
solutions
to (1.1) that are globally analytic on the entire
$xt$-plane and that decay exponentially as $x\to\pm\infty$
at each fixed $t\in\bold R.$
We derive an explicit formula, namely (4.11), and
its equivalents (4.12), (5.14), and (6.9), in order
to write such solutions in a compact form
utilizing matrix exponentials. These solutions can alternatively
be expressed explicitly as algebraic combinations of exponential,
trigonometric, and polynomial functions
of $x$ and $t.$ We also present an explicit formula, namely
(5.6), and its equivalents (6.14) and (6.15), for the magnitude of such solutions.

The idea behind our method is similar to that used in [10] to
generate exact solutions to the Korteweg-de Vries equation on the
half line, and we are motivated by the use of the IST with rational
scattering data. This involves representing the corresponding
scattering data in terms of a matrix realization [11], establishing
the separability of the kernel of a related Marchenko integral
equation by expressing that kernel in terms of a matrix exponential,
solving the Marchenko integral equation algebraically, and observing
that the procedure leads to exact solutions to the NLS equation even
when the input to the Marchenko equation does not necessarily come
from any scattering data.

For the general use of rational scattering data in inverse
scattering theory, the reader is referred, for example, to [8,9,17]
and the references therein.

Our method has several advantages:

\item{(i)} It is generalizable to obtain similar
explicit formulas for exact solutions to other integrable nonlinear
PDEs where the IST involves the use of a Marchenko integral
equation. For example, a similar method has been used [10] for the
half-line Korteweg-de Vries equation, and it can be applied to other
equations such as the defocusing nonlinear Schr\"odinger equation,
the modified Korteweg-de Vries equation, and the sine-Gordon
equation.

\item{(ii)} It is generalizable to the matrix versions of
the aforementioned integrable nonlinear PDEs. For example, a similar
method has been applied in the second author's Ph.D. thesis [20] to
the matrix NLS equation in the focusing case with a cubic nonlinearity.

\item{(iii)} As seen from our explicit formula
(4.11), our exact solutions are represented in a simple and compact
form in terms of a square matrix $A,$ a constant row vector $C,$ and
a constant column vector $B,$ where $A$ appears in a matrix
exponential. Such matrix exponentials can be ``unpacked" in a
straightforward way to express our exact solutions in terms of
exponential, trigonometric, and polynomial functions. Depending on
the size of $A,$ such unpacked expressions may take many pages to
display. Our explicit formula and its equivalents allow easy
evaluation of such unpacked expressions and numerical evaluations on
such exact solutions, as evident from the examples in available
Mathematica notebooks [39].

\item{(iv)} Our method easily deals with nonsimple bound-state poles
and the time evolution of the corresponding bound-state norming
constants. In the literature, nonsimple bound-state poles are
usually avoided due to mathematical complications. We refer the
reader to [32], where nonsimple bound-state poles are investigated
and complications are encountered. A systematic treatment of
nonsimple bound states has recently been given [13].

\item{(v)} Our method might be generalizable to the case
where the matrix $A$ becomes
a linear operator on a separable Hilbert space. Such a generalization
on which we are currently
working would allow us to solve
the NLS equation with initial potentials more general than those
considered in our paper.

Our method to produce exact solutions to the NLS equation is based
on using the IST [1-3,5,31,37]. There are also other methods to
obtain solutions to (1.1). Such methods include the use of a Darboux
transformation [16], the use of a B\"acklund transformation [12,14],
the bilinear method of Hirota [28], the use of various other
transformations such as the Hasimoto transformation [15,27], and
various other techniques [6] based on guessing the form of a
solution and adjusting various parameters. The main idea behind
using the transformations of Darboux and B\"acklund is to produce
new solutions to (1.1) from previously known solutions, and other
transformations are used to produce solutions to the NLS equation
from solutions to other integrable PDEs. The basic idea behind the
method of Hirota is to represent the solution as a ratio of two
functions and to determine those two functions by solving some
corresponding coupled differential equations. A unified treatment of
Hirota's method, the IST, and B\"acklund transformation to obtain
soliton solutions with simple and multiple poles for the Sine-Gordon
equation was given by P\"oppe by using Fredholm determinants [33].
Other techniques may use an ansatz such as determining $\Theta(x,t)$
and $M(x,t)$ by using $u(x,t)=e^{i\Theta(x,t)}M(x,t)$ in (1.1). For
example, trying
$$u(x,t)=e^{i(k_1x+k_2t+k_3)}f(k_4 x+k_5 t+k_6),\tag 1.2$$
where $k_j$ are constant real parameters
and $f$
is a real-valued smooth function, we get an exact solution
if we choose $k_2=1-k_1^2,$
$k_4=\pm 1,$ $k_5=\mp 2k_1,$ and $f$ as the hyperbolic
secant. One can also use the fact that if
$U(x,t)$ is a solution to (1.1), so is $e^{ic(x-ct)}U(x-2ct,t)$ for
any real constant $c.$ Multiplying a solution by a complex constant
of unit amplitude yields another solution, and hence such a phase factor
can always be omitted from the solution.

There are many references in which some exact solutions to (1.1) are
presented. For example, [38] lists five explicit solutions, one is
of the form of (1.2) with a constant $f,$ the second and third with
$f$ as the hyperbolic secant (those are one-soliton solutions with
simple poles), the fourth being periodic in $x,$ and the fifth is
the $n$-soliton solution. Another solution, which is periodic in
$x,$ is [6]
$$u(x,t)=ae^{2ia^2t}\left[
\displaystyle\frac{2b^2\cosh(2a^2b\sqrt{2-b^2}\,t)+2ib\sqrt{2-b^2}\,
\sinh(2a^2b\sqrt{2-b^2}\,t)}
{2\cosh(2a^2b\sqrt{2-b^2}\,t)-\sqrt{2}\sqrt{2-b^2}\cos(\sqrt{2}\,abx)}-1\right],\tag 1.3$$
where $a$ and $b$ are arbitrary real parameters. By letting $b\to 0$ in (1.3)
we get the solution
$$u(x,t)=a\,e^{2ia^2t}\displaystyle\frac{3+16ia^2t-16a^4t^2-4a^2x^2}{1+16a^4t^2+4a^2x^2}.$$
Another exact solution which is periodic in $x$ is presented [7] in
terms of the Jacobi elliptic functions. An exact solution to (1.1)
is displayed [22] in the form of a specific matrix realization and
is shown to be valid for $t\in[0,\epsilon)$ for some small
$\epsilon$ and $x\in[0,+\infty).$ In their celebrated paper [37]
Zakharov and Shabat list the one- and $n$-soliton solutions as well
as a one-soliton solution with a double pole, which is obtained from
a two-soliton solution with simple poles by letting those poles
coalesce. In [32] solitons with multiple eigenvalues are analyzed
and a one-soliton solution with a double pole and a one-soliton
solution with a triple pole are listed with the help of the symbolic
software REDUCE, by stating that ``in an actual calculation it is
very complex to exceed" higher order poles. With our method in this
paper we show that such solitons with any number of poles and any
multiplicities can be easily expressed by using an appropriate
representation. Let us also add that some periodic or almost
periodic solutions can be obtained in terms of two hyperelliptic
theta functions [29,30], and the scattering data for (2.1) can be
constructed corresponding to certain initial profiles [34,35].

In order to appreciate the power of our method, to see why it
produces new solutions, and to understand why it produces exact
solutions that are either impossible or difficult to produce by
other methods, let us consider the following. When the matrix size
is large (imagine $A$ being a $1000\times 1000$ matrix) we have an
explicit compact formula for an exact solution as in (4.11) or its
equivalents (4.12), (5.14), and (6.9). By using a computer algebra
system we can explicitly express such a solution in terms of
exponential, trigonometric, and polynomial functions of $x$ and $t$
(even though such an expression will take thousands of pages to
display, we are able to write such an expression thanks to our
explicit formula). The only explicit formula in the literature
comparable to ours is the formula for the $n$-soliton solution
without multiplicities. Our own explicit formula yields that
explicit $n$-soliton solution without multiplicities in a trivial
case; namely, when $A$ is a diagonal matrix of distinct entries with
positive real parts, as indicated in (7.1). Our explicit formula
also easily yields the $n$-soliton solution with arbitrary
multiplicities as a special case. Dealing with even a single soliton
with multiplicities has not been an easy task in other methods; for
example, the exact solution example presented in [37] for a
one-soliton solution with a double pole, which is obtained by
coalescing two distinct poles into one, contains a typographical
error, as pointed out in [32].

Our paper is organized as follows. In Section 2 we present the
preliminaries and outline the Marchenko method to solve the inverse
scattering problem for the Zakharov-Shabat system given in (2.1),
summarize the IST for the NLS equation, and list in (2.12) the time
evolution of the norming constants in a compact form [13], which is
valid even when bound-state poles may have multiplicities greater
than one. In Section 3 we consider (2.1) with some rational
scattering data, which in turn we express in terms of the matrices
$A,$ $B,$ $C$ given in (3.5)-(3.7), respectively. In Section~4, we
derive the explicit formula (4.11) for our exact solutions $u(x,t)$
to (1.1) in terms of $A,$ $B,$ $C,$ and we show that such solutions
have analytic extensions to the entire $xt$-plane when the real
parts of the eigenvalues of $A$ are positive. In Section 5 we
independently and directly verify that (4.11) is a solution to (1.1)
as long as the matrix $\Gamma(x;t)$ given in (4.7) is invertible,
which is assured on the entire $xt$-plane when the real parts of the
eigenvalues of $A$ are positive. In Section~5 we also show that
$|u(x,t)|^2$ can be expressed in terms of the logarithmic derivative
of the determinant of $\Gamma(x;t).$ In Section 6 we remove the
positivity restriction on the real parts of the eigenvalues of $A,$
and we enlarge the class of exact solutions represented by our
explicit formula (4.11) or its equivalents (4.12), (5.14), and
(6.9). Finally, in Section 7 we present some examples showing how
our explicit formula easily yields exact solutions to (1.1)
expressed in terms of exponential, trigonometric, and polynomial
functions, and we also mention the availability of various
Mathematica notebooks [39], in which the user can easily modify the
input and produce various exact solutions to (1.1) and their
animations by specifying $A,$ $B,$ $C.$

\vskip 10 pt
\noindent {\bf 2. PRELIMINARIES}
\vskip 3 pt

Consider the Zakharov-Shabat system on the full line
$$\bmatrix \xi\\
\noalign{\medskip}
\eta\endbmatrix'=\bmatrix -i\lambda&q(x)\\
\noalign{\medskip}
-\overline{q(x)}&i\lambda\endbmatrix
\bmatrix \xi\\
\noalign{\medskip}
\eta\endbmatrix,\qquad x\in\bold R,\tag 2.1$$
where the prime denotes the $x$-derivative,
$\lambda$ is the complex-valued spectral parameter,
$q$ is a complex-valued integrable potential, and the bar denotes
complex conjugation. There are two linearly independent vector
solutions to (2.1)
denoted by
$\psi(\lambda,x)$ and $\phi(\lambda,x),$ which are usually known as
the Jost solutions and are uniquely obtained by
imposing the respective asymptotic
conditions
$$\psi(\lambda,x)=\bmatrix 0\\
\noalign{\medskip}
e^{i\lambda x}\endbmatrix+o(1),\qquad x\to+\infty,\tag 2.2$$
$$\phi(\lambda,x)=\bmatrix e^{-i\lambda x}\\
\noalign{\medskip}
0\endbmatrix +o(1),\qquad x\to-\infty.$$
The transmission coefficient $T,$ the left reflection coefficient
$L,$ and the right reflection coefficient $R$ are then obtained through the
asymptotics
$$\psi(\lambda,x)=
\bmatrix e^{-i\lambda x} L(\lambda)/T(\lambda)\\
\noalign{\medskip} e^{i\lambda x}/T(\lambda)\endbmatrix
+o(1)
,\qquad x\to-\infty,\tag 2.3$$
$$\phi(\lambda,x)=
\bmatrix e^{-i\lambda x}/T(\lambda)\\
\noalign{\medskip} e^{i\lambda x}R(\lambda)/T(\lambda)\endbmatrix
+o(1) ,\qquad x\to+\infty.\tag 2.4$$ For further information on
these scattering solutions to (2.1) we refer the reader to
[1-3,5,31,37] and the references therein.

Besides scattering solutions to (2.1), we have so-called bound-state
solutions, which are square-integrable solutions to (2.1). They
occur at the poles of $T$ in the upper half complex plane $\bold
C^+.$ Let us denote the (distinct) bound-state poles of $T$ by
$\lambda_j$ for $j=m+1,\dots,m+n,$ and suppose that the multiplicity
of the pole at $\lambda_j$ is given by $n_j.$ The reason to start
indexing the bound states with $j=m+1$ instead of $j=1$ is for
notational convenience. It is known [1-3,5,31,37] that there is only
one linearly independent square-integrable vector solution to (2.1)
when $\lambda=\lambda_j$ for $j=m+1,\dots,m+n.$ Associated with each
such $\lambda_j,$ we have $n_j$ bound-state norming constants
$c_{js}$ for $s=0,\dots,n_j-1.$

The inverse scattering problem for (2.1) consists of recovery of
$q(x)$ for $x\in\bold R$ from an appropriate set of scattering data
such as the one consisting of the reflection coefficient
$R(\lambda)$ for $\lambda\in\bold R$ and the bound-state information
$\{\lambda_j,\{c_{js}\}_{s=0}^{n_j-1}\}_{j=m+1}^{m+n}.$ This problem
can be solved via the Marchenko method as follows [1-3,5,31,37]:

\item{a)} From the scattering data $\{R(\lambda),\{\lambda_j\},\{c_{js}\}\},$ form the
Marchenko kernel $\Omega$ as
$$\Omega(y):=\displaystyle\frac{1}{2\pi}\displaystyle\int_{-\infty}^\infty d\lambda\,
R(\lambda)\,e^{i\lambda y}+\sum_{j=m+1}^{m+n}\sum_{s=0}^{n_j-1} c_{js}
\displaystyle\frac{y^s}{s!}\,e^{i\lambda_j y}.
\tag 2.5$$

\item{b)} Solve the Marchenko equation
$$K(x,y)-\overline{\Omega(x+y)}+\int_x^\infty dz\int_x^\infty ds\, K(x,s)\,
\Omega(s+z)\,\overline{\Omega(z+y)}=0,\quad y>x.\tag 2.6$$

\item{c)} Recover the potential $q$ from the solution $K(x,y)$ to the
Marchenko equation via
$$q(x)=-2 K(x,x).\tag 2.7$$

\item{d)} Having determined $K(x,y),$ also determine
$$G(x,y):=-\int_x^\infty dz\, \overline{K(x,z)}\,\overline{\Omega(z+y)}.\tag 2.8$$
Then, obtain the Jost solution
$\psi(\lambda,x)$ to the Zakharov-Shabat system
(2.1)-(2.2) via
$$\psi(\lambda,x)=\bmatrix 0\\
\noalign{\medskip}
e^{i\lambda x}\endbmatrix+\int_x^\infty dy\, \bmatrix K(x,y)\\
\noalign{\medskip} G(x,y)\endbmatrix e^{i\lambda y}.\tag 2.9$$ Note
that $|q(x)|^2$ can be calculated from (2.7) or equivalently by
using [37]
$$\int_x^\infty dz\,|q(z)|^2=-2\,G(x,x),\quad
|q(x)|^2=2\,\displaystyle\frac{d G(x,x)}{dx}.\tag 2.10$$

The initial-value problem for (1.1) consists of recovery of $u(x,t)$
for $t>0$ when $u(x,0)$ is available. When $u(x,0)=q(x),$ where $q$
is the potential appearing in (2.1), it is known that such an
initial-value problem can be solved [1-3,5,31,37] by the method of
IST as indicated in the following diagram:

$$\CD u(x,0)
@>\text{direct scattering}>> \{R(\lambda),\{\lambda_j\},\{c_{js}\}\}  \\
@V{\text{solution to NLS}}VV @VV{\text{time evolution}}V \\
u(x,t) @<<\text{inverse scattering}< \{R(\lambda;t),\{\lambda_j\},\{c_{js}(t)\}\}
\endCD$$

\noindent The application of the IST
involves three steps:

\item{i)} Corresponding to the initial
potential $q(x),$ obtain the scattering data at $t=0;$ namely,
the reflection coefficient
$R(\lambda),$ the bound-state poles $\lambda_j$ of $T(\lambda),$ and
the norming constants $c_{js}.$

\item{ii)} Let the initial scattering data evolve in time.
The time-evolved reflection coefficient
$R(\lambda;t)$ is obtained from the reflection coefficient
$R(\lambda)$ via
$$R(\lambda;t)=R(\lambda)\,e^{4i\lambda^2 t}.\tag 2.11$$
The bound-state poles $\lambda_j$ and $T(\lambda)$ do not change in time.
The time evolution of the bound-state norming constants
$c_{js}(t)$ has been known when $s=0$ as
$$c_{j0}(t)=c_{j0}\,e^{4i\lambda_j^2 t},\qquad
j=n+1,\dots,m+n.$$ The time evolution of the remaining terms has
recently been analyzed in a systematic way [13], and the evolution
of $c_{js}(t)$ is described by the product of $e^{4i\lambda_j^2 t}$
and a polynomial in $t$ of order $s;$ we have [13]
$$\bmatrix c_{j(n_j-1)}(t)& \dots
& c_{j 0}(t)\endbmatrix= \bmatrix c_{j(n_j-1)}& \dots & c_{j
0}\endbmatrix e^{-4i A_j^2t},\tag 2.12$$ where $A_j$ is the matrix
defined in (3.3). See also [32], where a more complicated procedure
is given to obtain $c_{j s}(t).$

\item{iii)} Solve the inverse scattering problem for (2.1) with the
time-evolved scattering data $\{R(\lambda;t),\{\lambda_j,
\{c_{js}(t)\}_{s=0}^{n_j-1}\}_{j=m+1}^{m+n}\}$ in order to
obtain the time-evolved potential. It turns out that the
resulting time-evolved
potential $u(x,t)$ is a solution to (1.1) and reduces to $q(x)$ at $t=0.$
This inverse problem can be solved by the Marchenko method as outlined in Section~4
by replacing the kernel $\Omega(y)$ with its time-evolved version
$\Omega(y;t),$ which is obtained by replacing in (2.5)
$R(\lambda)$ by $R(\lambda;t)$ and
$c_{js}$ by $c_{js}(t).$

\vskip 10 pt
\noindent {\bf 3. REPRESENTATION OF THE SCATTERING DATA}
\vskip 3 pt

We are interested in obtaining explicit solutions to (1.1) when the
reflection coefficient $R(\lambda)$ appearing in (2.4) is a rational
function of $\lambda$ with poles occurring in $\bold C^+.$ For this
purpose we will use a method similar to the one developed in [10]
and already applied to the half-line Korteweg-de Vries equation. We
will first represent our scattering data in terms of a constant
square matrix $A,$ a constant column vector $B,$ and a constant row
vector $C.$ We will then rewrite the Marchenko kernel $\Omega(y)$
given in (2.5) in terms of $A,$ $B,$ $C.$ It will turn out that the
time-evolved kernel $\Omega(y;t)$ will be related to $\Omega(y)$ in
an easy manner. By solving the Marchenko equation (2.6) with the
time-evolved kernel $\Omega(y;t),$ we will obtain the time-evolved
solution $K(x,y;t),$ from which we will recover the time-evolved
potential $u(x,t)$ in a manner analogous to (2.7).

In this section we show how to construct $A,$ $B,$ $C$ from
some rational scattering data associated with the Zakharov-Shabat system. We show that
our exact solutions can be obtained
by choosing our triplet $A,$ $B,$ $C$ as in (3.5)-(3.7), where
$\lambda_j$ are distinct and $c_{j(n_j-1)}\ne 0$ for
$j=1,\dots,m+n.$

When the rational $R(\lambda)$ has poles at
$\lambda_j$ in $\bold C^+$
with multiplicity $n_j$ for $j=1,\dots,m,$ since
$R(\lambda)\to 0$ as $\lambda\to \infty,$ the partial fraction expansion
of $R(\lambda)$ can be written as
$$R(\lambda)=\sum_{j=1}^{m}
\sum_{s=1}^{n_j}\displaystyle\frac{(-i)^s r_{js}}{(\lambda-\lambda_j)^s},\tag 3.1$$
for some complex coefficients $r_{js}.$
Note that we can represent the inner summation in (3.1) in the form
$$\sum_{s=1}^{n_j}\displaystyle\frac{(-i)^s r_{js}}{(\lambda-\lambda_j)^s}
=-iC_j(\lambda-iA_j)^{-1}B_j,\tag 3.2$$
where, for $j=1,\dots,m,$ we have defined
$$A_j:=\bmatrix -i\lambda_j&
-1&0&\dots&0&0\\
0& -i\lambda_j& -1&\dots&0&0\\
0&0&-i\lambda_j& \dots&0&0\\
\vdots&\vdots &\vdots &\ddots&\vdots&\vdots\\
0&0&0&\dots&-i\lambda_j&-1\\
0&0&0&\dots&0&-i\lambda_j\endbmatrix,
\quad
B_j:=\bmatrix
0\\
0\\
0\\
\vdots\\
0\\
1\endbmatrix,\tag 3.3$$
$$C_j:=\bmatrix r_{j n_j}& \dots
&
r_{j 1}\endbmatrix,$$
so that
$$\lambda-iA_j=\bmatrix \lambda-\lambda_j&
i&0&\dots&0&0\\
0& \lambda-\lambda_j& i&\dots&0&0\\
0&0&\lambda-\lambda_j& \dots&0&0\\
\vdots&\vdots &\vdots &\ddots&\vdots&\vdots\\
0&0&0&\dots&\lambda-\lambda_j&i\\
0&0&0&\dots&0&\lambda-\lambda_j\endbmatrix,$$
$$
(\lambda-iA_j)^{-1}
=\bmatrix \frac{1}{\lambda-\lambda_j}&
\frac{-i}{(\lambda-\lambda_j)^2}&\frac{(-i)^2}{(\lambda-\lambda_j)^3}&\dots&
\frac{(-i)^{n_j-2}}{(\lambda-\lambda_j)^{n_j-1}}&\frac{(-i)^{n_j-1}}
{(\lambda-\lambda_j)^{n_j}}\\
0& \frac{1}{\lambda-\lambda_j}& \frac{-i}{(\lambda-\lambda_j)^2}&\dots&
\frac{(-i)^{n_j-3}}{(\lambda-\lambda_j)^{n_j-2}}&
\frac{(-i)^{n_j-2}}{(\lambda-\lambda_j)^{n_j-1}}\\
0&0&\frac{1}{\lambda-\lambda_j}& \dots&
\frac{(-i)^{n_j-4}}{(\lambda-\lambda_j)^{n_j-3}}&\frac{(-i)^{n_j-3}}
{(\lambda-\lambda_j)^{n_j-2}}\\
\vdots&\vdots &\vdots &\ddots&\vdots&\vdots\\
0&0&0&\dots&\frac{1}{\lambda-\lambda_j}&\frac{-i}{(\lambda-\lambda_j)^2}\\
0&0&0&\dots&0&\frac{1}{\lambda-\lambda_j}\endbmatrix.$$
We remark that the row vector $C_j$ contains $n_j$ entries, the column vector
$B_j$ contains $n_j$ entries, and $A_j$ is an $n_j\times n_j$ square matrix,
$(-A_j)$ is in a Jordan canonical form,
and that $(\lambda-iA_j)^{-1}$ is an upper triangular Toeplitz matrix.

As for the bound states, for $j=m+1,\dots,m+n,$ let us use (3.3) to define
the $n_j\times n_j$ matrix $A_j$ and
the column $n_j$-vector $B_j,$ and let $C_j$ be
the row $n_j$-vector defined as
$$C_j:=\bmatrix c_{j(n_j-1)}& \dots
&
c_{j 0}\endbmatrix,$$
so that the summation term in (2.5) is obtained as
$$\sum_{s=0}^{n_j-1} c_{js}
\displaystyle\frac{y^s}{s!}\,e^{i\lambda_j y}
=-\displaystyle\frac{i}{2\pi}\displaystyle\int_{-\infty}^\infty
d\lambda\,C_j(\lambda-iA_j)^{-1}B_je^{i\lambda y},
\qquad y>0.\tag 3.4$$

Now let us define the $p\times p$ block diagonal matrix $A$ as
$$A:=\bmatrix A_1&0&\dots&0\\
0&A_2&\dots&0\\
0&0& \dots&0\\
\vdots&\vdots &\ddots&\vdots\\
0&0&\dots&A_{m+n}\endbmatrix,\tag 3.5$$
where $p$ is the integer given by
$$p:=\sum_{j=1}^{m+n}n_j.$$
Similarly, let us define the column $p$-vector $B$ as
$$B:=\bmatrix B_1\\
B_2\\
\vdots\\
B_{m+n}\endbmatrix,\tag 3.6$$
and the row $p$-vector $C$ as
$$C:=\bmatrix C_1&C_2&\dots&C_{m+n}\endbmatrix.\tag 3.7$$

Without loss of generality we can assume that $\lambda_j$ for $j=1,\dots,m+n$
are all distinct;
in case one of $\lambda_j$ for $j=1,\dots,m$ coincides
with one of $\lambda_j$ for $j=m+1,\dots,m+n,$ we can simply combine the corresponding
blocks in (3.5) to reduce the number of blocks in $A$ by one. In case more such
$\lambda_j$ coincide, we can proceed in a similar way so that each block
in (3.5) will be associated with a distinct
$\lambda_j.$
Similarly, we can combine the corresponding
blocks in each of (3.6) and (3.9) so that the sizes of $B$ and $C$ will
be compatible with the size of $A.$

Consider the function $P(\lambda)$ defined as
$$P(\lambda):=-iC(\lambda-iA)^{-1}B,\qquad \lambda\in\bold C,\tag 3.8$$
with the triplet $A,$ $B,$ $C,$ where the constant matrices $A,$
$B,$ $C$ have sizes $p\times p,$ $p\times 1,$ and $1\times p,$
respectively, and the singularities of $P(\lambda)$ occur at the
eigenvalues of $iA.$ Such a representation is called {\it minimal}
[11] if there do not exist constant matrices $\tilde A,$ $\tilde B,$
$\tilde C$ with sizes $\tilde p\times\tilde p,$ $\tilde p\times 1,$
and $1\times\tilde p,$ respectively, such that $P(\lambda)=-i\tilde
C(\lambda-i\tilde A)^{-1}\tilde B$ and $\tilde p<p.$ There always
exists a triplet corresponding to a minimal representation. It is
known [11] that the realization with the triplet $A,$ $B,$ $C$ is
minimal if and only if the two $p\times p$ matrices defined as
$$\text{col}_p(C,A):=\bmatrix C\\
CA\\
\vdots
\\
CA^{p-1}\endbmatrix,\quad
\text{row}_p(A,B):=\bmatrix B&AB&\dots&A^{p-1}B\endbmatrix,\tag 3.9$$
both have rank $p.$

The following theorem shows that, for the sake of constructing exact
solutions to (1.1), it is sufficient to consider
only the triplet $A,$ $B,$ $C$ given in
(3.5)-(3.7) with distinct
$\lambda_j$ for $j=1,\dots,m+n$ because any other
triplet $\tilde A,$ $\tilde B,$ $\tilde C$
with sizes $p\times p,$ $p\times 1,$ and $1\times p,$ respectively,
can be equivalently expressed in terms of $A,$ $B,$ $C.$

\noindent{\bf Theorem 3.1} {\it
Given any arbitrary triplet $\tilde A,$ $\tilde B,$ $\tilde C$
with sizes $p\times p,$ $p\times 1,$ and $1\times p,$ respectively,
there exists a triplet $A,$ $B,$ $C$ having the form given in
(3.5)-(3.7), respectively, which yields the same
exact solution to (1.1). The construction of $A,$ $B,$ $C$
can be achieved by using}
$$\tilde A=MAM^{-1},\quad \tilde B=MSB,\quad C=\tilde CMS,\tag 3.10$$
{\it where $M$ is an invertible matrix whose columns consist of
the generalized eigenvectors of $\tilde A,$ the matrix $S$ is an
upper triangular Toeplitz matrix
commuting with $A,$ and the complex entries of $C$ are chosen
as in (3.10).}

\noindent PROOF: Since $(-A)$ is in the Jordan canonical
form, any given $\tilde A$ can be converted to $A$ by using
$\tilde A=MAM^{-1},$ where $M$ is a matrix
whose columns are formed by using the generalized
eigenvectors of $(-\tilde A).$ Next, consider all matrices $S$ commuting with
$A.$ Any such matrix has the block diagonal form
$$S:=\bmatrix S_1&0&\dots&0\\
0&S_2&\dots&0\\
0&0& \dots&0\\
\vdots&\vdots &\ddots&\vdots\\
0&0&\dots&S_{m+n}\endbmatrix,
\quad S_j:=
\bmatrix \alpha_{jn_j}&\alpha_{j(n_j-1)}&\dots&\alpha_{j1}\\
0&\alpha_{jn_j}&\dots&\alpha_{j2}\\
0&0& \dots&\alpha_{j3}\\
\vdots&\vdots &\ddots&\vdots\\
0&0&\dots&\alpha_{jn_j}\endbmatrix,
\tag 3.11$$
where $n_j$ is the order of the pole $\lambda_j$
for $j=1,\dots,m+n,$ and the constants $\alpha_{js}$ are arbitrary.
We will determine such $\alpha_{js}$ and hence $S$
itself by using $M^{-1}\tilde B=SB.$ Note that $SB$ is the column $p$-vector
consisting of $m+n$ column blocks, where the $j$th block has entries
$\alpha_{j1},\dots,\alpha_{jn_j}.$ Thus, $S$ is unambiguously constructed
 from $M$ and $\tilde B.$ Having constructed
$M$ and $S$ from $\tilde A$ and $\tilde B,$ we finally choose the
complex entries in the matrix $C$ appearing in (3.7) so that
$C=\tilde CMS.$ Let us now show the equivalence of the representation
with the triplet $\tilde A,$ $\tilde B,$ $\tilde C$ and that with the triplet
$A,$ $B,$ $C.$ From (3.8) we see that we must show
$$-iC(\lambda-iA)^{-1}B=-i\tilde C(\lambda-i\tilde A)^{-1}\tilde B.\tag 3.12$$
Since $SA=AS$ and $MA=\tilde AM,$ we also have
$$S(\lambda-iA)^{-1}=(\lambda-iA)^{-1}S,\quad
M(\lambda-iA)^{-1}=(\lambda-i\tilde A)^{-1}M.\tag 3.13$$
Replacing $C$ by $\tilde CMS$ on the left hand side of (3.12) and using
(3.13), we establish the equality in (3.12).
Similarly, replacing $C$ by $\tilde CMS$ on the right
hand side of (4.2) and using
$MA=\tilde AM$ and $SA=AS$ and (3.13), we prove that
$\Omega(y;t)$ remains unchanged if $A,$ $B,$ $C$ are replaced with
$\tilde A,$ $\tilde B,$ $\tilde C,$ respectively, in (4.2).
Hence the triplet $A,$ $B,$ $C$ and the triplet
$\tilde A,$ $\tilde B,$ $\tilde C$ yield the same solution
to (1.1). \qed

Note that the invertibility
of $S$ is not needed in Theorem~3.1. On the other hand, from (3.11)
it is seen that $S$ is invertible if and only if
$\alpha_{jn_j}\ne 0$ for $j=1,\dots,m+n.$
In the rest of this section we will give a characterization for the
minimality of the representation in (3.8) with the triplet
$A,$ $B,$ $C$ given in
(3.5)-(3.7). We will show that
as long as $\lambda_j$ are distinct and
$c_{j(n_j-1)}\ne 0$ in (3.7) for $j=1,\dots,m+n,$
the triplet $A,$ $B,$ $C$ given in
(3.5)-(3.7) can be used to recover in the
form of (4.11)
our exact solutions to (1.1). First, we need a result needed in the
proof of Theorem~3.3.

\noindent{\bf Proposition 3.2} {\it The matrix $\text{row}_p(A,B)$
defined in (3.9) is invertible if and only if
$\lambda_j$ for $j=1,\dots,m+n$ appearing in (3.5) are distinct.}

\noindent PROOF: It is enough to prove that the rows of
$\text{row}_p(A,B)$ are linearly independent
if and only if
$\lambda_j$ for $j=1,\dots,m+n$ are distinct.
We will give the proof by showing that a row-echelon
equivalent matrix $T$ defined below
has linearly independent rows.
Using (3.5) and (3.6) we get
$$\text{row}_p(A,B)=\bmatrix
\text{row}_p(A_1,B_1)\\
\noalign{\medskip}
\text{row}_p(A_2,B_2)\\
\vdots\\
\text{row}_p(A_{m+n},B_{m+n})\endbmatrix.$$
With the help of (3.3) we see that the $n_j\times p$ matrix
$\text{row}_p(A_j,B_j)$ is given by
$$\bmatrix 0&0&0&\dots&0&(-1)^{p-1}\\
\noalign{\medskip}
0&0&0&\dots&(-1)^{p-2}&(-1)^{p-1}(p-1)(i\lambda_j)\\
\vdots&\vdots&\vdots&\adots&\vdots&\vdots\\
0&-1&2i\lambda_j&\dots&(-1)^{p-2}(p-2)(i\lambda_j)^{p-3}&(-1)^{p-1}(p-1)(i\lambda_j)^{p-2}\\
\noalign{\medskip}
1&-i\lambda_j&(i\lambda_j)^2&\dots&(-1)^{p-2}(i\lambda_j)^{p-2}&(-1)^{p-1}(i\lambda_j)^{p-1}
\endbmatrix,$$
where we observe the binomial
expansion of $(-i\lambda_j-1)^s$ in the $(s-1)$st column.
Put $\sigma(k):=\#\{j:\ n_j\ge k\},$ i.e.
the number of Jordan blocks of $A$ of size at least $k.$
Then, $m+n=\sigma(1)\ge \sigma(2)\ge \sigma(3)\ge ....$ By reordering
the rows of $\text{row}_p(A,B)$ we obtain a
row-equivalent $p\times p$ echelon
matrix $T$ such that $T_{r1}=0$ for
$r>\sigma(1),$ $T_{r2}=0$ for
$r>\sigma(1)+\sigma(2),$
$T_{r3}=0$ for
$r>\sigma(1)+\sigma(2)+\sigma(3),$ etc.,
while the submatrices consisting of the elements
$T_{rs}$ for
$r=\sigma(1)+\dots+\sigma(k-1)+1,\dots,
\sigma(1)+\dots+\sigma(k)$
and $s=k,k+1,\dots,p$
have the form
$$\bmatrix
1&a_{k1}\mu_1&a_{k2}\mu_1^2&\dots&a_{k(k-1)}\mu_1^{p-k-1}\\
\vdots&\vdots&\vdots&\ddots&\vdots\\
1&a_{k1}\mu_{\sigma(k)}&a_{k2}\mu_{\sigma(k)}^2&\dots&a_{k(k-1)}\mu_{\sigma(k)}^{p-k-1}\endbmatrix,
\tag 3.14$$
where apart from a sign,
the coefficients
$a_{ks}$ are the binomial coefficients and hence nonzero,
and the constants
$\mu_1,\dots,\mu_{\sigma(k)}$ correspond to a rearrangement
of those of $-i\lambda_j$ for which
$n_j\ge k.$ Since
the matrix given in (3.14) can be written as the product of a Vandermonde
matrix and a nonsingular diagonal matrix, its rows are
linearly independent if and only if $\lambda_j$ with
$n_j\ge k$ are distinct.
 From the echelon structure of the matrix $T$
it then follows that all the rows of $T,$ and hence the rows
of $\text{row}_p(A,B)$ are linearly independent. \qed

\noindent{\bf Theorem 3.3} {\it The triplet $A,$ $B,$ $C$ given in
(3.5)-(3.7) corresponds to a minimal representation in (3.8)
if and only if $\lambda_j$ are all distinct and
$c_{j(n_j-1)}\ne 0$ for $j=1,\dots,m+n.$}

\noindent PROOF: Note that the matrix $S$ defined in (3.11) commute with
$A,$ and we have $SA=AS$ and $S_jA_j=A_jS_j$ for $j=1,\dots,m+n.$
Let us use a particular choice for $S_j$ by letting
$\alpha_{j1}=c_{j0},$ $\alpha_{j2}=c_{j1},\dots,
\alpha_{jn_j}=c_{j(n_j-1)}.$ Thus, $S$ is invertible if and only if
$c_{j(n_j-1)}\ne 0$ for $j=1,\dots,m+n.$ Let us define
the column $p$-vector $\hat B$ and the row $p$-vector $\hat C$ via
$\hat B=SB$ and $\hat CS=C.$ As in the proof of (3.12) in
Theorem~3.1 we obtain
$$-iC(\lambda-iA)^{-1}B=-i\hat C(\lambda-iA)^{-1}\hat B.$$
and hence the representation in (3.8) with the triplet
$A,$ $B,$ $C$ is equivalent to that with
$A,$ $\hat B,$ $\hat C.$ From the statement containing
(3.9) it then follows that our theorem is proved if we can show that
$\text{row}_p(A,\hat B)$ and $\text{col}_p(\hat C,A)$ are both invertible
if and only if $\lambda_j$ are all distinct and
$c_{j(n_j-1)}\ne 0$ for $j=1,\dots,m+n.$
Below we will prove that $\text{row}_p(A,\hat B)$ and $\text{col}_p(\hat C,A)$
are invertible if and only if $\text{row}_p(A,B)$ and $S$ are invertible.
Our theorem then follows from Proposition~3.2 and the fact that
$S$ is invertible if and only if $c_{j(n_j-1)}\ne 0$ for $j=1,\dots,m+n.$
Since $SA=AS$ and $SB=\hat B,$ from (3.9) we obtain
$$S\, \text{row}_p(A,B)=\text{row}_p(A,SB)=\text{row}_p(A,\hat B)
,$$
and hence $\text{row}_p(A,\hat B)$ is invertible if and only
if $\text{row}_p(A,B)$ and $S$ are invertible.
We complete the proof by showing that
$\text{col}_p(\hat C,A)$
is invertible if and only if $\text{row}_p(A,B)$ is invertible.
Define the $n_j\times n_j$ matrix $J_j$ and
the $p\times p$ matrix $J$ as
$$J_j:=\bmatrix 0&0&\dots&0&1\\
0&0&\dots&1&0\\
\vdots&\vdots&\ddots&\vdots&\vdots\\
0&1&\dots&0&0\\
1&0&\dots&0&0
\endbmatrix,\quad
J:=\bmatrix J_1&0&\dots&0\\
0&J_2&\dots&0\\
0&0& \dots&0\\
\vdots&\vdots &\ddots&\vdots\\
0&0&\dots&J_{m+n}\endbmatrix,\tag 3.15$$
where $1$ appears along the trailing diagonal
of $J_j.$ Let us use the superscript $T$ to denote the matrix transpose.
Note that
$$J_j^{-1}=J_j, \quad J_j^T=J_j, \quad
J^{-1}=J,\quad J^T=J.$$
It can be verified from (3.5) that
$JAJ=A^T.$ Using (3.6), (3.7), and (3.15), since
$\hat C=CS^{-1}$ we get
$\hat C=B^TJ.$ Thus, we have
$$\left(\text{col}_p(\hat C,A)\right)^T=
\text{row}_p(A^T,\hat C^T)
=\text{row}_p(A^T,JB)=J\,\text{row}_p(A,B).$$
Since $J$ is invertible, our proof is complete. \qed

\vskip 10 pt
\noindent {\bf 4. EXPLICIT SOLUTIONS TO THE NLS EQUATION}
\vskip 3 pt

In the previous section we have constructed $A,$ $B,$ $C$ given
in (3.5)-(3.7), respectively,
from some rational scattering data of the
Zakharov-Shabat system.
In this section we solve the corresponding
time-evolved Marchenko equation explicitly for $x\ge 0$
in terms of such $A,$ $B,$ $C.$ Such solutions lead to explicit
solutions to (1.1) via the formula given in (4.11). We then show that
such solutions have analytic extensions to the entire $xt$-plane if
the real parts of the eigenvalues of
$A$ are positive, which is equivalent to having $\lambda_j\in\bold C^+$ for
$j=1,\dots,m+n$ in (3.3). We also analyze various properties of the
key matrices $Q(x;t),$ $N(x),$ and $\Gamma(x;t)$ that appear in
(4.7)-(4.9) and
that are used to construct
our exact solutions.

For $y\ge 0,$ with the help (3.2), (3.4), and a contour integration along the
boundary of $\bold C^+,$ we evaluate the kernel $\Omega(y)$ defined in (2.5) as
$$\Omega(y)=Ce^{-Ay}B,\qquad y\ge 0.\tag 4.1$$
Note that (4.1) yields a separable kernel
for the Marchenko integral equation in (2.6) because from
$$\Omega(x+y)=Ce^{-Ax}e^{-Ay}B,$$
we see that $\Omega(x+y)$ is the Euclidean product of the
row $p$-vector $Ce^{-Ax}$ and the column $p$-vector $e^{-Ay}B.$
As a result of this separability we are able to solve the Marchenko integral
equation (2.6) exactly by algebraic means.

At this point we discuss the time evolution of the scattering data in
more detail. Using (2.11) we can express the time-evolved Marchenko integral
kernel as
$$\Omega(y;t)=\displaystyle
\frac{1}{2\pi}\int_{-\infty}^\infty d\lambda\,R(\lambda)\,
e^{4i\lambda^2t}e^{i\lambda y}+\sum_{j=m+1}^{m+n}\sum_{s=0}^{n_j-1}c_{js}(t)\,
\displaystyle
\frac{y^s}{s!}\,e^{i\lambda_jy},$$
where $c_{js}(t)$ satisfies (2.12). This time-evolved
kernel is seen to satisfy the first order PDE
$$\Omega_t(y;t)+4i\Omega_{yy}(y;t)=0,$$
provided the integral $\int_{-\infty}^\infty
d\lambda\,(1+\lambda^2)\,|R(\lambda)|$ exists. Such PDEs for
Marchenko kernels have been studied in [4] for a variety of
nonlinear evolution equations and in [20] for the matrix NLS
equation. Here we use (4.1) as an initial condition in solving this
PDE and write
$$\Omega(y;t)=Ce^{-Ay-4iA^2t}B,\qquad y\ge 0.\tag 4.2$$
In other words, $\Omega(y;t)$ is obtained from $\Omega(y)$ by replacing $C$
in (4.1) by $Ce^{-4iA^2t}.$
Let us use a dagger to denote the matrix adjoint (complex conjugate
and transpose). Since $\Omega(y;t)$ is a scalar, its complex conjugate
is the same as its adjoint and we have
$$\Omega(y;t)^\dagger=B^\dagger e^{-A^\dagger y+4i(A^\dagger)^2t}C^\dagger.\tag 4.3$$
Comparing with (2.6) we obtain the time-evolved
Marchenko integral equation as
$$K(x,y;t)-\Omega(x+y;t)^\dagger+\int_x^\infty dz\int_x^\infty ds\, K(x,s;t)\,
\Omega(s+z;t)\,\Omega(z+y;t)^\dagger=0,\quad y>x.\tag 4.4$$
Using (4.2) and (4.3) in (4.4), we see that we can look for a solution in the form
$$K(x,y;t)=H(x;t)\,e^{-A^\dagger y+4i(A^\dagger)^2t}C^\dagger,\tag 4.5$$
where $H(x;t)$ is to be determined. Using (4.5) in (4.4), we obtain
$$H(x;t)\,\Gamma(x;t)=B^\dagger e^{-A^\dagger x},\tag 4.6$$
where we have defined
$$\Gamma(x;t):=I+Q(x;t)\,N(x),\tag 4.7$$
with $I$ denoting the $p\times p$ identity matrix and
$$Q(x;t):=\int_x^\infty ds\,e^{-A^\dagger s+4i(A^\dagger)^2t}C^\dagger
Ce^{-As-4iA^2t},\tag 4.8$$
$$N(x):=\int_x^\infty dz\,e^{-A z}BB^\dagger e^{-A^\dagger z}.\tag 4.9$$

Using (4.6) in (4.5) we can write the solution to (4.4) as
$$K(x,y;t)=B^\dagger e^{-A^\dagger x}\Gamma(x;t)^{-1}
e^{-A^\dagger y+4i(A^\dagger)^2t}C^\dagger,\tag 4.10$$
provided $\Gamma(x;t)$ is invertible. We will prove
the invertibility of $\Gamma(x;t)$ in Theorem 4.2.
In analogy to (2.7) we get the time-evolved potential as
$u(x,t)=-2\,K(x,x;t),$ and hence
the solution to (1.1) is obtained as
$$u(x,t)=-2B^\dagger e^{-A^\dagger x}\Gamma(x;t)^{-1}
e^{-A^\dagger x+4i(A^\dagger)^2t}C^\dagger.\tag 4.11$$ It is
possible [19] to write (4.11) as the ratio of two determinants as
$$u(x,t)=\displaystyle\frac{\det F(x;t)}{\det \Gamma(x;t)},\tag 4.12$$
where the $(p+1)\times(p+1)$ matrix $F(x;t)$ is given by
$$F(x;t):=\bmatrix 0&2B^\dagger e^{-A^\dagger x}\\
e^{-A^\dagger x+4i(A^\dagger)^2t}C^\dagger&\Gamma(x;t)\endbmatrix.$$

We end this section by listing some useful properties of the matrices
$Q(x;t),$ $N(x),$ and $\Gamma(x;t).$

\noindent{\bf Proposition 4.1} {\it The matrices $Q(x;t)$ and $N(x)$ defined in
(4.8) and (4.9), respectively, satisfy}
$$Q(x;t)=e^{-A^\dagger x+4i(A^\dagger)^2t}Q(0;0)\,
e^{-Ax-4iA^2t},\quad
N(x)=e^{-Ax}N(0)\,
e^{-A^\dagger x},\tag 4.13$$
{\it and the integrals in (4.8) and (4.9) converge
for all $x,t\in\bold R$ as long as
all the eigenvalues of
$A$ have positive real parts.}

\noindent PROOF: By replacing $s$ and $z$ with $s-x$ and $z-x$ in
(4.8) and (4.9), respectively, we obtain (4.13). From (4.8) and (4.9), we then get
$$Q(0;0)=\int_0^\infty ds\,[Ce^{-As}]^\dagger
[Ce^{-As}],\quad
N(0)=
\int_0^\infty dz\,[e^{-A z}B][e^{-A z}B]^\dagger.\tag 4.14$$
If $\epsilon>0$ is chosen such that the real parts of the eigenvalues of $A$ exceed
$\epsilon,$ then in any matrix norm $||\cdot||$ we have
$\left\|e^{-A z}\right\|=O(e^{-\epsilon z})$ and
$\left\|e^{-A^\dagger z}\right\|=O(e^{-\epsilon z})$ as $z\to+\infty.$
Hence, the integrals in (4.14) converge, and as a consequence of (4.13)
the integrals in (4.8) and (4.9) converge
for all $x,t\in\bold R.$
\qed

The next theorem shows that the matrix $\Gamma(x;t)$ defined in
(4.7) is invertible for all $x,t\in\bold R$ as long as the eigenvalues of
$A$ have positive real parts. In fact, in that case
$\Gamma(x;t)$ has a positive determinant for all $x,t\in\bold R.$

\noindent{\bf Theorem 4.2} {\it Assume that the eigenvalues of
$A$ have positive real parts.
Then, for every $x,t\in\bold R$ we have the following:}
\item{(i)} {\it The matrices $Q(x;t)$ and $N(x)$ defined in
(4.8) and (4.9), respectively, are positive and selfadjoint.
Consequently, there exist unique positive selfadjoint matrices
$Q(x;t)^{1/2}$ and $N(x)^{1/2}$ such that
$Q(x;t)=Q(x;t)^{1/2}Q(x;t)^{1/2}$ and $N(x)=N(x)^{1/2}N(x)^{1/2}.$}
\item{(ii)} {\it The matrix $\Gamma(x;t)$ defined in (4.7) is invertible.}
\item{(iii)} {\it The determinant of $\Gamma(x;t)$ is positive.}

\noindent PROOF: In our proof let us write $Q$ and $N$ for $Q(x;t)$
and $N(x),$ respectively. The positivity and selfadjointness of $Q$
and $N$ are a direct consequence of the fact that each of the
integrands in (4.8) and (4.9) can be written as the product of a
matrix and its adjoint; hence [23] we have proved (i). From the
Sherman-Morrison-Woodbury formula [23] it follows that
$$[I+Q^{1/2}(Q^{1/2}N)]^{-1}=I-Q^{1/2}[I+(Q^{1/2}N)Q^{1/2}]^{-1}Q^{1/2}N,$$
and hence $(I+QN)$ is invertible if and only if
$(I+Q^{1/2}NQ^{1/2})$ is invertible; on the
other hand, the latter can be written as
$[I+(Q^{1/2}N^{1/2})(Q^{1/2}N^{1/2})^\dagger]$
due to the selfadjointness of
$Q^{1/2}$ and $N^{1/2},$ and hence it is invertible,
establishing (ii). From the two matrix identities
$$\bmatrix I&0\\ Q^{1/2}N&I\endbmatrix \bmatrix I&Q^{1/2}
\\ -Q^{1/2}N&I\endbmatrix \bmatrix I&-Q^{1/2}
\\ 0&I\endbmatrix =\bmatrix I&0
\\ 0&I+Q^{1/2}NQ^{1/2}\endbmatrix,$$
$$\bmatrix I&-Q^{1/2}\\ 0&I\endbmatrix
\bmatrix I&Q^{1/2}\\ -Q^{1/2}N&I\endbmatrix
\bmatrix I&0\\ Q^{1/2}N&I\endbmatrix
=\bmatrix I+QN&0\\ 0&I\endbmatrix,$$
it follows that $I+QN$ and $(I+Q^{1/2}NQ^{1/2})$ have the same determinant.
Thus, we have (iii) as a result of
the fact that
the determinant of $[I+(Q^{1/2}N^{1/2})(Q^{1/2}N^{1/2})^\dagger]$ is positive. \qed

\noindent{\bf Proposition 4.3} {\it Assume that the eigenvalues of
$A$ have positive real parts. Then, for all $x,t\in\bold R$ the matrices $Q(x;t),$
$N(x),$ $\Gamma(x;t)$ defined in (4.7)-(4.9) satisfy:}
$$Q_x=-A^\dagger Q-QA,\quad N_x=-AN-NA^\dagger,\quad Q_t=4i[(A^\dagger)^2
Q-QA^2],\tag 4.15$$
$$\Gamma^\dagger=I+NQ,\quad
\Gamma^{-1}Q=Q(\Gamma^\dagger)^{-1},
\quad (\Gamma^\dagger)^{-1}N=N\Gamma^{-1}.\tag 4.16$$

\noindent PROOF: We obtain (4.15) from (4.13), or (4.8) and (4.9), through
differentiation. Using the selfadjointness of $Q$ and $N$ proved in
Theorem 4.2, from (4.7) we obtain (4.16). \qed

\noindent{\bf Theorem 4.4} {\it For every $x,t\in\bold R,$
the matrices $Q(x;t)$ and $N(x)$ defined in
(4.8) and (4.9), respectively, are simultaneously invertible
for all $x,t\in\bold R$
if and only if the realization in (4.1) of
$\Omega(y)$ with the triplet $A,$ $B,$ $C$ is minimal and the eigenvalues
of $A$ have positive real parts.}

\noindent PROOF: From (4.13) we see that it is enough
to prove that $Q(0;0)$ and $N(0)$ defined in (4.14) are invertible.
The integrals in (4.14)
are convergent as a result of
the positivity of the real parts of the eigenvalues of $A.$
If $Q(0;0)g=0$ for some vector $g\in\bold C^p,$
then from (4.14) we see that $C e^{-As}g=0$ for all $s\ge 0.$ By analytic
continuation this implies that $C e^{-As}g=0$ for all $s\in\bold C$ and
hence
$$CA^kg=0,\qquad k=0,1,\dots.\tag 4.17$$
Similarly, if $N(0)h=0$ for some vector $h\in\bold C^p,$ using
(4.14) we conclude that
$$B^\dagger (A^\dagger)^kh=0,\qquad k=0,1,\dots.\tag 4.18$$
It is known [11] that the realization in (3.8) or (4.1) for the
triplet $A,B,C$ is minimal if and only if the two matrices given in
(3.9) both have rank $p,$ where we recall that the size of $A$ is
$p\times p,$ that of $B$ is $p\times 1,$ and that of $C$ is $1\times
p.$ On the other hand, the ranks of the two matrices in (3.9) are
both $p$ if and only if (4.17) and (4.18) have only the trivial
solutions $g=0$ and $h=0,$ respectively. \qed

For any fixed $x_0\in\bold R,$ by shifting the dummy integration
variable in (4.9) we get
$$N(x)=e^{-A(x-x_0)}N(x_0)e^{-A^\dagger(x-x_0)},$$
and similarly from (4.8) for any $x_0,t_0\in\bold R$ we get
$$Q(x;t)=e^{-A^\dagger(x-x_0)+4i(A^\dagger)^2(t-t_0)}
Q(x_0;t_0)e^{-A(x-x_0)-4iA^2(t-t_0)}.$$
Thus, we have the following observations.

\noindent{\bf Corollary 4.5} {\it Assume that the eigenvalues of
$A$ have positive real parts. Then, the matrix $N(x)$
defined in (4.9) is invertible for all $x\in\bold R$
if and only if it is invertible at any one particular value of $x.$
Similarly, $Q(x;t)$ defined in (4.8) is invertible for all $x,t\in\bold R$ if and only if
it is invertible at any one particular point on the $xt$-plane.}

\noindent{\bf Proposition 4.6} {\it If the eigenvalues
of $A$ have positive real parts, then the matrix
$\Gamma(x;t)$ defined in (4.7) satisfies
$\Gamma(x;t)\to I$ as $x\to+\infty.$ Additionally, if $Q(0;0)$ and
$N(0)$ given in (4.14) are invertible, then
$\Gamma(x;t)^{-1}\to 0$ exponentially as $x\to-\infty,$ where
$I$ and $0$ are the $p\times p$ unit and zero matrices, respectively.}

\noindent PROOF: As stated in Proposition 4.1,
since the integrals in (4.8) and (4.9) converge,
$\Gamma(x;t)\to I$ as $x\to+\infty$ follows from (4.7)-(4.9). To obtain the limit
for $\Gamma(x;t)^{-1}$ as $x\to-\infty,$ let us first define
$$Y(x;t):=e^{A^\dagger x}\Gamma(x;t)\,e^{A^\dagger x}.\tag 4.19$$
Using (4.13) in (4.19) we get
$$Y(x;t)=Q(0;t)\,e^{-2Ax}N(0)\left[I+N(0)^{-1}e^{2Ax}Q(0;t)^{-1}e^{2A^\dagger x}\right].
\tag 4.20$$ Note that, from Theorem 4.2 it follows that $N(0)^{-1}$
and $e^{2Ax}Q(0;t)^{-1}e^{2A^\dagger x}$ are positive selfadjoint
matrices. Using the Sherman-Morrison-Woodbury formula [23] as in the
proof of Theorem 4.2, we see that the inverse of the matrix in the
brackets in (4.20) exists, and for all $x\in\bold R$ we have
$$Y(x;t)^{-1}=\left[I+N(0)^{-1}e^{2Ax}Q(0;t)^{-1}e^{2A^\dagger x}\right]^{-1}
N(0)^{-1}e^{2Ax}Q(0;t)^{-1}.\tag 4.21$$
Further, since the eigenvalues of
$A$ and $A^\dagger$ have strictly positive real parts, for each fixed $t\in\bold R$
we conclude, as in the proof of
Proposition 4.1, that there exists $\epsilon>0$
such that
$\left\|e^{Ax}\right\|=O(e^{\epsilon x})$ and
$\left\|e^{A^\dagger x}\right\|=O(e^{\epsilon x})$ as $x\to-\infty$
in any matrix norm $||\cdot||.$ Hence, from (4.21) we see that
$Y(x;t)^{-1}\to 0$ exponentially as $x\to-\infty,$
and writing (4.19) in the form
$$\Gamma(x;t)^{-1}=e^{A^\dagger x}Y(x;t)^{-1}\,e^{A^\dagger x},$$
we also see that $\Gamma(x;t)^{-1}\to 0$ exponentially as $x\to-\infty.$
\qed

\vskip 10 pt
\noindent {\bf 5. FURTHER PROPERTIES OF OUR EXPLICIT SOLUTIONS}
\vskip 3 pt

We have obtained certain explicit solutions to (1.1) in the form of (4.11)
by starting with some rational scattering data for (2.1)
and by constructing the corresponding matrices
$A,$ $B,$ and $C$ given in (3.5)-(3.7), respectively.
In this section we will show that (4.11) is a solution to
(1.1) no matter how the triplet $A,$ $B,$ $C$ is chosen, as long as the
matrix $\Gamma(x;t)$ defined in (4.7) is invertible.
For example, from Theorem~4.2 it follows that
$\Gamma(x;t)^{-1}$ exists on the entire
$xt$-plane and thus (4.11) is a solution to
(1.1) when the eigenvalues of
$A$ have positive real parts.

The purpose of this section is threefold. We will first
obtain some useful representations for $|u(x,t)|^2$ corresponding
to $u(x,t)$ given in (4.11). Next, we will prove that
$u(x,t)$ given in (4.11) is a solution to (1.1) as long as
$\Gamma(x;t)^{-1}$ exists. Then, we will consider further
properties of such solutions.

We can evaluate $|u(x,t)|^2$ from (4.11) directly. Alternatively,
we can recover it by using the time-evolved
analog of (2.10), namely
$$\int_x^\infty dz\,|u(z,t)|^2=-2\,G(x,x;t),\quad
|u(x,t)|^2=2\,\displaystyle\frac{\partial G(x,x;t)}{\partial x},\tag 5.1$$
where, in comparison with (2.8), we see that
$$G(x,y;t):=-\int_x^\infty dz\, \Omega(y+z;t)^\dagger
K(x,z;t)^\dagger.\tag 5.2$$
 From (4.3), (4.8), (4.10), and (5.2), we get
$$G(x,y;t)=-B^\dagger
e^{-A^\dagger y}\Gamma(x;t)^{-1}Q(x;t)e^{-A x}B.\tag 5.3$$
Using (5.3) in (5.1), with the help of (4.15), (4.16), and
$$(\Gamma^{-1})_x=-\Gamma^{-1}\Gamma_x\Gamma^{-1},
\quad (\Gamma^{-1})_t=-\Gamma^{-1}\Gamma_t\Gamma^{-1},\tag 5.4$$
we obtain
$$|u(x,t)|^2=4B^\dagger
e^{-A^\dagger x}\Gamma(x;t)^{-1}[A^\dagger Q(x;t)+Q(x;t) A]
[\Gamma(x;t)^\dagger]^{-1}e^{-A x}B.\tag 5.5$$

Next we show that $|u(x,t)|^2$ can be expressed in a simple form in terms of
the matrix $\Gamma(x;t)$ defined in (4.7). As indicated in Theorem 4.2, recall that
$\Gamma(x;t)$ has a positive determinant for all $x,t\in\bold R$
when the real parts of the eigenvalues of $A$ are positive.

\noindent{\bf Theorem 5.1} {\it The absolute square $|u(x,t)|^2$ of the solution to the
NLS equation can be written directly in terms of the determinant of
the matrix $\Gamma(x;t)$ defined in (4.7) so that}
$$|u(x,t)|^2=\displaystyle\frac{\partial}{\partial x}\left[
\displaystyle\frac{\partial\det\Gamma(x;t)/\partial x}{\det\Gamma(x;t)}\right]=
\displaystyle\frac{\partial^2}{\partial x^2}\left[\log \left(\det\Gamma(x;t)
\right)\right].\tag 5.6$$

\noindent PROOF: In terms of a matrix trace, from (5.1) and (5.3) we get
$$|u(x,t)|^2=-2\left[
B^\dagger
e^{-A^\dagger x}\Gamma^{-1}Q\,e^{-A x}B\right]_x=
2 \text{tr} \left[\Gamma^{-1}QN_x\right]_x,\tag 5.7$$
where we have used (4.9) and the fact that
in evaluating the trace of a product of two matrices
the order in the product can be changed.
With the help of (4.7), (4.15), (4.16), and the trace properties we obtain
$$\text{tr}\left[\Gamma^{-1}QN_x\right]=\text{tr}\left[-A-A^\dagger+(\Gamma^\dagger)^{-1}A+
\Gamma^{-1}A^\dagger\right],\tag 5.8$$
$$\text{tr}\left[\Gamma^{-1}Q_xN\right]=\text{tr}\left[-A-A^\dagger+(\Gamma^\dagger)^{-1}A+
\Gamma^{-1}A^\dagger\right].\tag 5.9$$
Thus, from (5.7)-(5.9) with the help of (4.7) we get
$$2 \text{tr} \left[\Gamma^{-1}QN_x\right]=
\text{tr} \left[\Gamma^{-1}Q_xN+\Gamma^{-1}QN_x\right]=
\text{tr} \left[\Gamma^{-1}\Gamma_x\right],$$
and hence
$$|u(x,t)|^2=\text{tr} \left[\Gamma^{-1}\Gamma_x\right]_x,$$
which can also be written as (5.6), as indicated in Theorem 7.3 on
p. 38 of [18]. \qed

We remark that (5.6) is a generalization of the formula given at the
end of Section~3 of [37], where the formula was obtained for the
$n$-soliton solution with simple poles. Thus, our method handles the
bound states with nonsimple poles easily even though nonsimple poles
have always caused complications in other methods and have mostly
been avoided in the literature.

Let us also remark that (1.1) has infinitely many conserved quantities expressed as trace formulas.
One such trace formula is given in the following.

\noindent {\bf Proposition 5.2} {\it
When the eigenvalues of
the matrix $A$ have positive real parts,
the function $u(x,t)$ given
in (4.11) satisfies the trace formula}
$$\displaystyle\int_{-\infty}^\infty dx\,|u(x,t)|^2=
\text{tr} \left[A+A^\dagger\right]=2\displaystyle\sum_{j=1}^{m+n} n_j\,\text{Im} [\lambda_j],
\tag 5.10$$
{\it where $\lambda_j$ and $n_j$ are the poles
in $\bold C^+$ and the corresponding multiplicities, respectively,
as in (3.3).}

\noindent PROOF: From (5.7) and (5.8) we see that
$$\displaystyle\int_{-\infty}^\infty dx\,|u(x,t)|^2=
\text{tr}\left[-A-A^\dagger+(\Gamma^\dagger)^{-1}A+
\Gamma^{-1}A^\dagger\right]\big|_{-\infty}^\infty.$$
As indicated in Proposition~4.6, we have
$\Gamma(x;t)\to I$ as $x\to+\infty$ and
$\Gamma(x;t)^{-1}\to 0$ as $x\to-\infty.$
Thus, we get the first equality in (5.10). Using (3.3) and (3.5), we can
write the trace of $(A+A^\dagger)$ in terms of the multiplicities and
imaginary parts of
$\lambda_j$ as indicated in the second equality in (5.10). \qed

\noindent {\bf Theorem 5.3} {\it
The function $u(x,t)$ given
in (4.11) satisfies (1.1) with any $p\times p$ matrix $A,$
column $p$-vector $B,$ and row $p$-vector $C$ as long as
the matrix $\Gamma(x;t)$ defined in (4.7) is invertible.
In particular, if all eigenvalues of $A$ have positive real parts, then
$u(x,t)$ given
in (4.11) satisfies (1.1) on the entire $xt$-plane.}

\noindent PROOF:
With the help of (4.15), (4.16), and (5.4),
through straightforward differentiation and after some
simplifications, from (4.11) we get
$$iu_t=8B^\dagger e^{-A^\dagger x}\Gamma^{-1}[
(A^\dagger)^2+
QA^2N]\Gamma^{-1}e^{-A^\dagger x+4i(A^\dagger)^2t}C^\dagger,\tag 5.11$$
$$u_x=4B^\dagger e^{-A^\dagger x}\Gamma^{-1}[
A^\dagger-QAN]\Gamma^{-1}e^{-A^\dagger x+4i(A^\dagger)^2t}C^\dagger,$$
$$\aligned
u_{xx}=8B^\dagger e^{-A^\dagger x}\Gamma^{-1}
[&(A^\dagger)^2-2QAN\Gamma^{-1}QAN+2A^\dagger
\Gamma^{-1}QAN
-2A^\dagger \Gamma^{-1} A^\dagger
\\
&+2QAN\Gamma^{-1} A^\dagger+QA^2N]
\Gamma^{-1}e^{-A^\dagger x+4i(A^\dagger)^2t}C^\dagger,\endaligned
\tag 5.12$$
$$2uu^\dagger u=-16B^\dagger e^{-A^\dagger x}\Gamma^{-1}
[(A^\dagger Q+QA)(\Gamma^\dagger)^{-1}
(AN+NA^\dagger)]\Gamma^{-1}e^{-A^\dagger x+4i(A^\dagger)^2t}C^\dagger.\tag 5.13$$
Using (4.16) and (5.11)-(5.13), and noting that
$u^\dagger=\overline{u},$ we verify that (1.1)
is satisfied. Let us note that (5.13) could also be obtained
directly by multiplying (4.11) and (5.5). \qed

\noindent {\bf Theorem 5.4} {\it Assume that the eigenvalues of
$A$ have positive real parts and that
the matrices
$Q(0;0)$ and $N(0)$ given in (4.14) are invertible, or
equivalently, assume that the representation in
(3.8) with the triplet $A,$ $B,$ $C$ is minimal and the eigenvalues
of $A$ have positive real parts.
Then, for each fixed $t\in\bold R$
the solution $u(x,t)$ given
in (4.11) vanishes exponentially as $x\to\pm\infty.$}

\noindent PROOF: From (4.11) and the fact that
$\Gamma(x;t)\to I$ as $x\to+\infty,$ it follows that
$u(x,t)\to 0$ exponentially as $x\to +\infty$ for each fixed $t\in\bold R.$
Let us write (4.11) as
$$u(x,t)=-2B^\dagger Y(x;t)^{-1}e^{4i(A^\dagger)^2t}C^\dagger,\tag 5.14$$
where $Y(x;t)$ is the matrix defined in (4.19). In the proof of Proposition~4.6, we
have shown that $Y(x;t)^{-1}\to 0$ exponentially
as $x\to-\infty.$ Hence, from (5.14) we can conclude
that for each fixed $t\in\bold R$ we
have $u(x,t)\to 0$ exponentially as $x\to-\infty.$ \qed

Let us remark that, if the eigenvalues
of $A$ have positive real parts,
when extended to the entire $x$-axis the solutions given in
(4.11) become multisoliton solutions, where the number of solitons,
multiplicity of the corresponding poles, and norming constants can be chosen at will.
This can also be seen by analytically continuing
the time-evolved Jost solution $\psi(\lambda,x;t)$ to the entire
$x$-axis,
by using (2.3), (2.9), and
$$\displaystyle\frac{L(\lambda;t)}{T(\lambda;t)}=\displaystyle\lim_{x\to-\infty}
\displaystyle\int_x^\infty dy\, K(x,y;t)\,e^{i\lambda (y-x)},\tag 5.15$$
by evaluating the integral with help of (4.10),
and by observing that the limit in (5.15) vanishes.

\vskip 10 pt
\noindent {\bf 6. GENERALIZATION}
\vskip 3 pt

In some parts of Sections 3-5 we have assumed that $\lambda_j$
values appearing in (3.3) and in the matrix $A$ given in (3.5)
are all located in $\bold C^+.$ In this section we relax that restriction and allow some
or all $\lambda_j$ to be located in the lower half complex plane
$\bold C^-.$ Our only restriction
will be that no $\lambda_j$ will be real and no two distinct $\lambda_j$
will be symmetrically located with respect to
the real axis in the complex plane. This restriction is mathematically
equivalent to the disjointness of
the sets $\{\lambda_j\}_{j=1}^{m+n}$
and $\{\overline\lambda_j\}_{j=1}^{m+n}.$
Under this restriction we will show that $u(x,t)$ given in (4.11) is
a solution to (1.1) in any region on the $xt$-plane
in which the matrix $\Gamma(x;t)$ defined in (4.7) is invertible. The only
change we need is that $Q(x;t)$ and $N(x)$ will no
longer be defined as in (4.8) and (4.9), but instead they will
be given as in (4.13), where we now let
$$Q(0;0)=\displaystyle\frac{1}{2\pi}\int_{\gamma}
d\lambda\,(\lambda+iA^\dagger)^{-1}C^\dagger C(\lambda-iA)^{-1},\tag 6.1$$
$$N(0)=\displaystyle\frac{1}{2\pi}\int_{\gamma}
d\lambda\,(\lambda-iA)^{-1}BB^\dagger(\lambda+iA^\dagger)^{-1},\tag 6.2$$
with $\gamma$ being any positively oriented simple closed contour enclosing
all $\lambda_j$ in such a way that all $\overline\lambda_j$
lie outside $\gamma.$

As the following proposition shows, the quantities
given in (6.1) and (6.2) are the
unique (selfadjoint) solutions to the respective
Lyapunov equations
$$Q(0;0)\,A+A^\dagger Q(0;0)=C^\dagger C,\tag 6.3$$
$$A\,N(0)+N(0)\,A^\dagger=BB^\dagger.\tag 6.4$$
We note that, using (4.13), we could also
write (6.3) and (6.4) in the equivalent form
$$Q(x;t)\,A+A^\dagger Q(x;t)=
e^{-A^\dagger x+4i(A^\dagger)^2t}C^\dagger C
e^{-A x-4iA^2t},\tag 6.5$$
$$A\,N(x)+N(x)\,A^\dagger=e^{-Ax}BB^\dagger
e^{-A^\dagger x}.\tag 6.6$$

\noindent {\bf Proposition 6.1} {\it Assume that none of the eigenvalues of
$A$ are purely imaginary and that no two
eigenvalues of $A$ are symmetrically located with
respect to the imaginary axis. Equivalently, assume that
$\{\lambda_j\}_{j=1}^{m+n}$ and $\{\overline\lambda_j\}_{j=1}^{m+n}$
are disjoint,
where $\lambda_j$ are the complex constants appearing in (3.3) and (3.5).
We then have the following:}
\item{(i)} {\it The matrix equations
given in (6.3) and (6.4) are each uniquely solvable.}
\item{(ii)} {\it The unique solutions $Q(0;0)$ and $N(0)$ are selfadjoint
matrices.}
\item{(iii)} {\it The unique solutions are given by
(6.1) and (6.2), respectively.}

\noindent PROOF: Note that (i) and (iii) directly follow from
Theorem 4.1 in Section I.4 of [21]. It is straightforward to show
that the adjoint of any solution to (6.3) or (6.4) is also a
solution to the same equation, and hence the unique solutions
$Q(0;0)$ and $N(0)$ must be selfadjoint. \qed

Next, without requiring that
all $\lambda_j$ appearing in (3.5) be located in
$\bold C^+,$ we will prove that
the matrix $u(x,t)$ given in (4.11) is a solution to
(1.1) as long as $\Gamma(x;t)$ defined in (4.7) is invertible. First,
we will write (4.11) in a slightly different but
equivalent form. Define
$$\Lambda(x;t):=I+P(x;t)^\dagger\,Q(0;0)\,P(x;t)\,N(0),\quad
P(x;t):=e^{-2Ax-4iA^2t}.\tag 6.7$$
Note that $\Gamma(x;t)$ is invertible if and only if
$\Lambda(x;t)$ is invertible because, by using (4.7), (4.13), and (6.7), we see that
$$\Gamma(x;t)=e^{A^\dagger x}\Lambda(x;t)\,e^{-A^\dagger x}.\tag 6.8$$
With the help of (6.8) we can write (4.11) in the equivalent form
$$u(x,t)=-2B^\dagger \Lambda(x;t)^{-1}P(x;t)^\dagger C^\dagger.\tag 6.9$$

\noindent {\bf Theorem 6.2} {\it Assume that
none of the eigenvalues of the matrix
$A$ in (3.5) are purely imaginary and that no two
eigenvalues of $A$ are symmetrically located with
respect to the imaginary axis. Equivalently, assume that
$\{\lambda_j\}_{j=1}^{m+n}$ and $\{\overline\lambda_j\}_{j=1}^{m+n}$
are disjoint,
where $\lambda_j$ are the complex constants appearing in (3.3) and (3.5).
Then, the quantity $u(x,t)$ given in (4.11), or equivalently in any of (4.12),
(5.14), and
(6.9) is a solution to (1.1) in any region
of the $xt$-plane where the matrix $\Lambda(x;t)$ defined in (6.7)
or equivalently
the matrix $\Gamma(x;t)$ given in (4.7) is invertible.}

\noindent PROOF: In our proof
let us write $u,$ $\Lambda,$ $P,$ $Q,$ $N$ for
$u(x,t),$ $\Lambda(x;t),$ $P(x;t),$ $Q(0;0),$ $N(0),$ respectively.
Without explicitly
mentioning it, we will use the selfadjointness $Q^\dagger=Q$ and
$N^\dagger=N$ established
in Proposition 6.1 as well as the fact that $P$ is invertible.
Proceeding as in the proof of Theorem~5.3,
using straightforward differentiation on (6.9) and after some
simplification we obtain
$$iu_t=8B^\dagger \Lambda^{-1}[(A^\dagger)^2
+P^\dagger QA^2PN]\Lambda^{-1}
P^\dagger C^\dagger,\tag 6.10$$
where we have used the fact that
$$\Lambda=I+P^\dagger QPN,\quad
(\Lambda^{-1})_t=-\Lambda^{-1} \Lambda_t\Lambda^{-1},\quad
P_t=-4iA^2P,\quad AP=PA.\tag 6.11$$
Similarly, by using (6.11) and
$$P_x=-2AP,\quad
(\Lambda^{-1})_x=-\Lambda^{-1} \Lambda_x\Lambda^{-1},$$
after some simplifications we obtain
$$u_x=4B^\dagger \Lambda^{-1}[A^\dagger
-P^\dagger QAPN]\Lambda^{-1}
P^\dagger C^\dagger,$$
$$\aligned
u_{xx}=8B^\dagger \Lambda^{-1}&
\left[(A^\dagger)^2-2A^\dagger
\Lambda^{-1} A^\dagger+
P^\dagger
QA^2PN+2A^\dagger
\Lambda^{-1} P^\dagger QAPN\right. \\
&  \left. +2P^\dagger
QAPN\Lambda^{-1} A^\dagger-
2P^\dagger QAPN \Lambda^{-1}P^\dagger QAPN
\right]\Lambda^{-1}
P^\dagger C^\dagger.\endaligned\tag 6.12$$
Next, with the help of (6.3) and (6.4) and
using $|u|^2u=uu^\dagger u,$ we obtain
$$\aligned 2|u|^2u=-16
B^\dagger \Lambda^{-1}
&\left[
P^\dagger QAP(\Lambda^\dagger)^{-1}AN+
P^\dagger QAP(\Lambda^\dagger)^{-1} NA^\dagger
\right. \\
&\left.
+P^\dagger A^\dagger QP
(\Lambda^\dagger)^{-1}AN
+P^\dagger A^\dagger QP(\Lambda^\dagger)^{-1}NA^\dagger\right]
\Lambda^{-1}
P^\dagger C^\dagger.\endaligned\tag 6.13$$
We see that (1.1) is satisfied, which is verified
by adding (6.10), (6.12), and (6.13) side by side and by using
$$QPN=(P^\dagger)^{-1}(\Lambda-I),\quad
(\Lambda^\dagger)^{-1}N=N \Lambda^{-1},\quad
NP^\dagger Q=(\Lambda^\dagger-I)P^{-1},$$
which directly follows from (6.7)
and the selfadjointness of $Q$ and $N.$ \qed

As the next theorem shows, if we remove the restriction
$\lambda_j\in\bold C^+$ then
the result in Theorem 5.1 still remains valid
in any region in the $xt$-plane where $\Gamma(x;t)$ or
equivalently $\Lambda(x;t)$ is invertible.

\noindent {\bf Theorem 6.3} {\it Assume that none of
the eigenvalues of the matrix
$A$ in (3.5) are purely imaginary and that no two
eigenvalues of $A$ are symmetrically located with
respect to the imaginary axis. Equivalently, assume that
$\{\lambda_j\}_{j=1}^{m+n}$ and $\{\overline\lambda_j\}_{j=1}^{m+n}$
are disjoint,
where $\lambda_j$ are the complex constants appearing in (3.3) and (3.5).
Then, in any region
of the $xt$-plane where the matrix $\Lambda(x;t)$ defined in (6.7)
or equivalently
the matrix $\Gamma(x;t)$ given in (4.7) is invertible,
the solution $u(x,t)$ given in (4.11) or equivalently
in (6.9) satisfies (5.6) or equivalently}
$$|u(x,t)|^2=\text{tr}\left[\displaystyle\frac
{\partial}{\partial x}
\left(\Gamma(x,t)^{-1}\displaystyle\frac{\partial \Gamma(x;t)}{\partial x}
\right)\right]=
\displaystyle\frac{\partial}{\partial x}\left[
\displaystyle\frac{\partial\det\Gamma(x;t)/\partial x}{\det\Gamma(x;t)}\right],\tag 6.14$$
$$|u(x,t)|^2=\text{tr}\left[\displaystyle\frac
{\partial}{\partial x}
\left(\Lambda(x,t)^{-1}\displaystyle\frac{\partial \Lambda(x;t)}{\partial x}
\right)\right]=
\displaystyle\frac{\partial}{\partial x}\left[
\displaystyle\frac{\partial\det\Lambda(x;t)/\partial x}{\det\Lambda(x;t)}\right].\tag 6.15$$

\noindent PROOF: Let us write $u,$ $\Lambda,$ $P,$ $Q,$ $N$ for
$u(x,t),$ $\Lambda(x;t),$ $P(x;t),$ $Q(x;t),$ $N(x),$ respectively.
Using the fact that, in evaluating the trace of a product
of two matrices we can change the order in the matrix product, from
(6.8) we obtain
$$\text{tr}\left[ \Gamma^{-1}\Gamma_x\right]=
\text{tr}\left[ \Lambda^{-1}\Lambda_x\right],$$
and hence it is sufficient to prove only (6.14).
From
(4.13) it follows that (6.5) and (6.6) are
equivalent to
the first two equations, respectively, in (4.15).
Note that (4.16) is still valid and is a direct consequence of
(4.7) and the selfadjointness of $Q$ and $N.$
Proceeding as in the proof of Theorem 5.1, with the help of (4.15),
(4.16), and (5.4) we obtain
$$\text{tr}\left[\Gamma^{-1}\Gamma_x\right]=
2\text{tr}\left[-A-A^\dagger+(\Gamma^\dagger)^{-1}A+
\Gamma^{-1}A^\dagger\right],$$
$$\aligned
\text{tr}\left[\Gamma^{-1}\Gamma_x\right]_x=
4\text{tr}&\left[\Gamma^{-1}(A^\dagger)^2+
(\Gamma^\dagger)^{-1}A^2-
\Gamma^{-1}A^\dagger\Gamma^{-1} A^\dagger
\right. \\
&\left. -(\Gamma^\dagger)^{-1}A (\Gamma^\dagger)^{-1}A
+2\Gamma^{-1}QAN\Gamma^{-1}A^\dagger\right].
\endaligned\tag 6.16$$
On the other hand, using the fact that
$|u|^2=uu^\dagger,$ from (4.11) we obtain
$$|u|^2=4\text{tr}\left[(AN+NA^\dagger)
\Gamma^{-1} (QA+A^\dagger Q)(\Gamma^\dagger)^{-1}\right],\tag 6.17$$
where we have also used (6.5) and (6.6). Using (4.16) and the
aforementioned property of the matrix trace, we can simplify the
right hand side of (6.17) and show that it is equal to the right
hand side of (6.16). Finally, as indicated in the proof of Theorem
5.1, the second equalities in (6.14) and (6.15) follow from Theorem
7.3 on p. 38 of [18]. \qed

\vskip 10 pt
\noindent {\bf 7. EXAMPLES}
\vskip 3 pt

Specific examples of our exact solutions can be obtained from the
explicit formula (4.11), or equivalently from any one of (4.12),
(5.14), and (6.9), by specifying $A,$ $B,$ and $C,$ where
$\Gamma(x;t)$ is the matrix defined in (4.7). We have made available
various Mathematica notebooks [39] in which the user can easily
perform the following steps and display the corresponding exact
solution $u(x,t)$ explicitly in terms of exponential, trigonometric,
and polynomial functions, verify that the resulting $u(x,t)$
satisfies (1.1), and animate $|u(x,t)|.$

\item{(i)} Input the matrices $A,$ $B,$ $C.$

\item{(ii)} Evaluate the matrix $\Gamma(x;t)$ as in (4.7), where
$Q(x;t)$ and
$N(x)$ are the matrices appearing in (4.13). In case all the eigenvalues of
$A$ lie in the right half complex plane,
evaluate $Q(0;0)$ and
$N(0)$ explicitly as
in (4.14) with the help of {\tt MatrixExp}, which is
used to evaluate matrix exponentials
in Mathematica.
In case some or all eigenvalues of
$A$ lie in the left half complex plane, use
(6.1) and (6.2) instead in order to evaluate explicitly
$Q(0;0)$ and
$N(0),$ respectively.

\item{(iii)} Having obtained $\Gamma(x;t),$ use (4.11)
or one of its equivalents (4.12),
(5.14), and (6.9) to display $u(x,t)$ explicitly
in terms of exponential, trigonometric, and polynomial
functions.

\item{(iv)} Using (5.6) or (4.11), evaluate $|u(x,t)|^2$ exactly and animate
$|u(x,t)|.$

\item{(v)} As an option, evaluate the quantities
$iu_t,$ $u_{xx},$ and $2|u|^2u,$ and verify directly that (1.1) is satisfied.

\noindent {\bf Example 7.1} The well-known ``$n$-soliton" to (1.1)
is obtained
when
$R(\lambda)\equiv 0$ and $T(\lambda)$ has
$n$ simple bound-state poles
in $\bold C^+.$
In this case, from (3.5)-(3.7) we see that $A,$ $B,$ and $C$ are given by
$$A=\bmatrix -i\lambda_1&0&\dots&0\\
0&-i\lambda_2&\dots&0
\\
\vdots&\vdots&\ddots&\vdots\\
0&0&\dots&-i\lambda_n\endbmatrix,\quad
B=\bmatrix 1\\
1\\
\vdots
\\ 1\endbmatrix,\quad
C=\bmatrix c_1&c_2&\dots&c_n\endbmatrix,\tag 7.1$$
where $\lambda_j$ are distinct and all lie in $\bold C^+.$
Using (4.7)-(4.9), the $(\alpha,\beta)$-entries of the matrices
$Q(x;t),$ $N(x),$ and $\Gamma(x;t)$ are easily evaluated as
$$N_{\alpha\beta}=\displaystyle\frac{ie^{i(\lambda_\alpha-\overline{\lambda}_\beta)x}}
{\lambda_\alpha-\overline{\lambda}_\beta}
,\quad
Q_{\alpha\beta}=\displaystyle\frac{i\overline{c}_\alpha c_\beta\,
e^{i(\lambda_\beta-\overline{\lambda}_\alpha)x+4i
(\lambda_\beta^2-\overline{\lambda}_\alpha^2)t}}{\lambda_\beta-\overline{\lambda}_\alpha}
,$$
$$\Gamma_{\alpha\beta}=
\delta_{\alpha\beta}-\displaystyle
\sum_{\gamma=1}^n\displaystyle\frac {\overline{c}_\alpha c_\gamma\,
e^{i(2\lambda_\gamma-\overline{\lambda}_\alpha-\overline{\lambda}_\beta)x
+4i(\lambda_\gamma^2-\overline{\lambda}_\alpha^2)t}}
{(\lambda_\gamma-\overline{\lambda}_\alpha)
(\lambda_\gamma-\overline{\lambda}_\beta)},$$ where
$\delta_{\alpha\beta}$ is the Kronecker delta. A Mathematica
notebook [39] is available, where the user can specify $n$ and
$\{\lambda_j,c_j\}_{j=1}^n$ and display the corresponding $u(x,t)$
explicitly in terms of exponential, trigonometric, and polynomial
functions and animate $|u(x,t)|.$

\noindent {\bf Example 7.2}
Choosing
$$A=\bmatrix 2&0\\
0&-1\endbmatrix,\quad
B=\bmatrix 1\\
1\endbmatrix,\quad
C=\bmatrix 1&-1\endbmatrix,$$
we evaluate $Q(0;0)$ and $N(0)$ using (6.3) and (6.4), respectively.
Then, with the help of (6.7) and (6.9) we obtain
$$u(x,t)=\displaystyle\frac{8 e^{4 i t}(9e^{-4 x}+16 e^{4 x})-
32e^{16i t}(4e^{-2 x}+9 e^{2 x})} {-128\cos (12 t)+4e^{-6 x}+16e^{6
x}+81e^{-2 x}+64e^{2 x}}.\tag 7.2$$ Note that one of the eigenvalues
of $A$ in this example is negative and the solution in (7.2) is not
a soliton solution. A Mathematica notebook containing the animation
of (7.2) is available [39].

\noindent {\bf Example 7.3}
Choosing
$$A=\bmatrix 2-i&-1\\
0&2-i\endbmatrix,\quad
B=\bmatrix 0\\
1\endbmatrix,\quad
C=\bmatrix 1+2i&-1+4i\endbmatrix,$$
we get $u(x,t)=\text{num}(x,t)/\text{den}(x,t),$ where
$$\aligned
\text{num}(x,t):=&
1024 e^{4(x+4 t)-2i(x-6t)}\left[
(12-9i)+100t+(5-10i)x\right]\\
&+131072e^{12(x+4 t)-2i(x-6t)}\left[
(1+4i)+(24+32i)t-(2-4i)x\right],\endaligned$$
$$\aligned\text{den}(x,t):=&25+65536
   e^{16 (4 t+x)}
   \\
   &+512 e^{8 (4 t+x)} \left[
   12800
   t^2+64 (20 x+43) t+160 x^2+304 x+207\right].
\endaligned$$
The solution in this example can be described as a soliton of double
multiplicity, and its Mathematica animation is available [39].

\noindent {\bf Example 7.4}
Choosing
$$A=\bmatrix 1&-1&0\\
0&1&-1\\
0&0&1\endbmatrix,\quad
B=\bmatrix 0\\
0\\
1\endbmatrix,\quad
C=\bmatrix 1&0&0\endbmatrix,$$
we easily obtain $u(x,t)=\text{num}(x,t)/\text{den}(x,t),$ where
$$\aligned
\text{num}(x,t)&:=32e^{-2(x-2it)}\left\{[-32768x^2+524288t^2+262144itx-65536it]
\right.\\ &+\left.e^{-4x}\left[90112t^2+15872x^2+131072t^2x^2+4096x^4
+196608xt^2\right.\right.\\ &+\left.\left.12288x^3+9216x+1344+1048576t^4
-32768itx^2-35840it\right.\right.\\ &-\left.\left.61440itx\right]
+e^{-8x}\left[128t^2-8x^2-24x-15-112it-64itx\right]\right\},\endaligned$$
$$\aligned\text{den}(x,t)&:=262144+e^{-4x}\left[262144x^4+589824x^2+393216x+524288x^3
\right.\\ &+\left.67108864t^4+8388608x^2t^2+122880\right]
+e^{-8x}\left[16384x^3+4096x^4\right.\\ &+\left.1048576t^4+15360x+344064t^2
+24576x^2+131072x^2t^2\right.\\ &+\left.393216xt^2+3648\right]+e^{-12x}.
\endaligned$$
The solution in this example can be described as a soliton of triple
multiplicity. A Mathematica notebook [39] is available for this
example and the corresponding animation.

In Figures 7.1-7.3 we present some snapshots of $|u(x,t)|$ appearing
in Examples 7.2-7.4. Further examples of exact solutions to (1.1)
expressed in terms of exponential, trigonometric, and polynomial
functions as well as their animations can be obtained with the help
of available Mathematica notebooks [39]. It can be directly verified
that $u(x,t)$ given in the above examples all satisfy (1.1). When
the matrix size for the $A,$ $B,$ $C$ becomes large, such
expressions become lengthy and yet can easily be displayed with the
help of Mathematica or any other symbolic software.

\vskip 10 pt

\noindent{\bf Acknowledgment}. The research leading to this article was
supported in part by the National Science Foundation under grant
DMS-0610494, the
Italian Ministry of Education and Research (MIUR) under PRIN grant no.
2006017542-003, and INdAM-GNCS.

\vskip 10 pt

\noindent {\bf REFERENCES}

\item{[1]} M. J. Ablowitz and P. A. Clarkson, {\it Solitons, nonlinear
evolution equations and inverse scattering,} Cambridge Univ. Press, Cambridge,
1991.

\item{[2]} M. J. Ablowitz, D. J. Kaup, A. C. Newell, and H. Segur, {\it The
inverse scattering transform-Fourier analysis for nonlinear problems,} Stud.
Appl. Math. {\bf 53}, 249--315 (1974).

\item{[3]} M. J. Ablowitz, B. Prinari, and A. D. Trubatch, {\it Discrete and
continuous nonlinear Schr\"odinger systems,}
Cambridge Univ. Press, Cambridge, 2004.

\item{[4]} M.J. Ablowitz, A. Ramani, and H. Segur, {\it A connection between
nonlinear evolution equations and ordinary differential equations of
$P$-type. II,} J. Math. Phys. {\bf 21}, 1006--1015 (1980).

\item{[5]} M. J. Ablowitz and H. Segur, {\it
Solitons and the inverse scattering
transform,} SIAM, Philadelphia, 1981.

\item{[6]} N. N. Akhmediev, V. M. Eleonskii, and N. E. Kulagin,
{\it First-order exact solutions of the nonlinear Schr\"odinger equation,}
Theoret. Math. Phys. {\bf 72}, 809--818 (1987).

\item{[7]} N. N. Akhmediev and V. I. Korneev,
{\it Modulation instability and periodic solutions of the
nonlinear Schr\"odinger equation,}
Theoret. Math. Phys. {\bf 69}, 1089--1093 (1986).

\item{[8]} T. Aktosun, M. H. Borkowski, A. J. Cramer, and L. C.
Pittman, {\it Inverse scattering with rational scattering
coefficients and wave propagation in nonhomogeneous media,} in: M. A.
Kaashoek, S. Seatzu, and C. van der Mee (eds.), {\it Recent advances
in operator theory and its applications,} Birkh\"auser, Basel, 2005,
pp. 1--20.

\item{[9]} T. Aktosun and M. Klaus, {\it Inverse theory: problem on
the line,} in: E. R. Pike and P. C. Sabatier (eds.), {\it
Scattering,} Academic Press, London, 2001, pp. 770--785.

\item{[10]} T. Aktosun and C. van der Mee, {\it Explicit solutions to the
Korteweg-de Vries equation on the half-line,} Inverse Problems {\bf 22},
2165--2174 (2006).

\item{[11]} H. Bart, I. Gohberg, and M. A. Kaashoek, {\it Minimal
factorization of matrix and operator functions,} Birkh\"auser,
Basel, 1979.

\item{[12]} M. Boiti and F. Pempinelli,
{\it Nonlinear Schr\"odinger equation, B\"acklund transformations and
Painlev\'e transcendents,}
Nuovo Cimento B (ser. 11) {\bf 59}, 40--58 (1980).

\item{[13]} T. Busse, Ph.D. thesis, University of Texas at Arlington,
in preparation.

\item{[14]} D. Cai and D. W. McLaughlin,
{\it Chaotic and turbulent behavior of unstable
one-dimensional nonlinear dispersive waves,}
J. Math. Phys. {\bf 41}, 4125--4153 (2000).

\item{[15]} A. Calini and T. Ivey, {\it Finite-gap solutions of the vortex
filament equation: genus one solutions and symmetric solutions,}
J. Nonlinear Sci. {\bf 15}, 321--361 (2005).

\item{[16]} R. C. Cascaval, F. Gesztesy, H. Holden,
and Y. Latushkin,
{\it Spectral analysis of Darboux transformations for the focusing NLS hierarchy,}
J. Anal. Math. {\bf 93}, 139--197 (2004).

\item{[17]} K. Chadan and P. C. Sabatier, {\it Inverse problems in quantum scattering
theory,} 2nd ed., Springer, New York, 1989.

\item{[18]} E. A. Coddington and N. Levinson, {\it Theory of ordinary
differential equations,} McGraw-Hill, New York, 1955.

\item{[19]} R. Courant and D. Hilbert, {\it Methods of mathematical physics,}
Vol. I, Interscience Publ., New York, 1989.

\item{[20]} F. Demontis, {\it Direct and inverse scattering of the matrix
Zakharov-Shabat system}, Ph.D. thesis, University of Cagliari, Italy, 2007.

\item{[21]} I. Gohberg, S. Goldberg, and
M. A. Kaashoek,
{\it Classes of linear operators,} Vol. I,
Birkh\"auser, Basel, 1990.

\item{[22]} I. Gohberg, M. A. Kaashoek, and A. L. Sakhnovich,
{\it Pseudo-canonical systems with rational Weyl functions: explicit formulas
and applications,} J. Differential Equations {\bf 146}, 375--398
(1998).

\item{[23]} G. H. Golub and C. F. Van Loan, {\it Matrix computations,}
3rd ed., Johns Hopkins Univ. Press, Baltimore, 1996.

\item{[24]} A. Hasegawa and M. Matsumoto, {\it Optical solitons in fibers,}
3rd ed., Springer, Berlin, 2002.

\item{[25]} A. Hasegawa and F. Tappert, {\it Transmission of stationary
nonlinear optical pulses in dispersive dielectric fibers. I. Anomalous
dispersion,} Appl. Phys. Lett. {\bf 23}, 142--144 (1973).

\item{[26]} A. Hasegawa and F. Tappert, {\it Transmission of stationary
nonlinear optical pulses in dispersive dielectric fibers. II. Normal
dispersion,} Appl. Phys. Lett. {\bf 23}, 171--172 (1973).

\item{[27]} R. Hasimoto, {\it A soliton on a vortex filament,} J. Fluid Mech.
{\bf 51}, 477--485 (1972).

\item{[28]} R. Hirota, {The direct method in soliton theory,}
Cambridge Univ. Press, Cambridge, 2004.

\item{[29]} A. R. Its and V. P. Kotljarov, {\it Explicit formulas for solutions of a
nonlinear Schr\"odinger equation,} Dokl. Akad. Nauk
Ukrain. SSR Ser. A {\bf 1976}, 965–-968, 1051 (1976) [Russian].

\item{[30]} A. M. Kamchatnov, {\it On improving the effectiveness
of periodic solutions of the NLS
and DNLS equations,} J. Phys. A {\bf 23}, 2945--2960 (1990).

\item{[31]} S. Novikov, S. V. Manakov, L. P. Pitaevskii, and V. E. Zakharov,
{\it Theory of solitons,}
Consultants Bureau, New York, 1984.

\item{[32]} E. Olmedilla, {\it Multiple pole solutions of the nonlinear
Schr\"odinger equation,} Phys. D {\bf 25}, 330--346 (1987).

\item{[33]} C. P\"oppe, {\it Construction of solutions of the sine-Gordon
equation by means of Fredholm determinants,} Phys. D {\bf 9}, 103--139 (1983).

\item{[34]} J. Satsuma and N. Yajima, {\it Initial value
problems of one-dimensional self-modulation of
nonlinear waves in dispersive media,}
Progr. Theoret. Phys. Suppl. {\bf 55}, 284--306 (1974).

\item{[35]} A. Tovbis and S. Venakidis, {\it The eigenvalue problem for the focusing
nonlinear Schr\"o\-din\-ger equation: new solvable cases,} Phys. D
{\bf 146}, 150--164 (2000).

\item{[36]} V. E. Zakharov, {\it Stability of periodic waves of
finite amplitude on the surface of a deep fluid,}
J. Appl. Mech. Tech. Phys. {\bf 4}, 190--194 (1968).

\item{[37]} V. E. Zakharov and A. B. Shabat, {\it Exact theory of
two-dimensional self-focusing and one-dimensional self-modulation of waves in
nonlinear media,} Sov. Phys. JETP {\bf 34}, 62--69 (1972).

\item{[38]} A. D. Polyanin and V. F. Zaitsev,
{\it Handbook of nonlinear partial differential equations,}
Chapman \& Hall/CRC Press, Boca Raton, 2004.

\item{[39]} http://omega.uta.edu/$\sim$aktosun

\newpage

\centerline{{\bf Fig.~7.1} Snapshots of $|u(x,t)|$
of Example 7.2 at $t=0.0,$ $0.1,$
$0.2,$ $0.3,$ $0.4,$ and $0.5.$}

\newpage

\psfig{figure=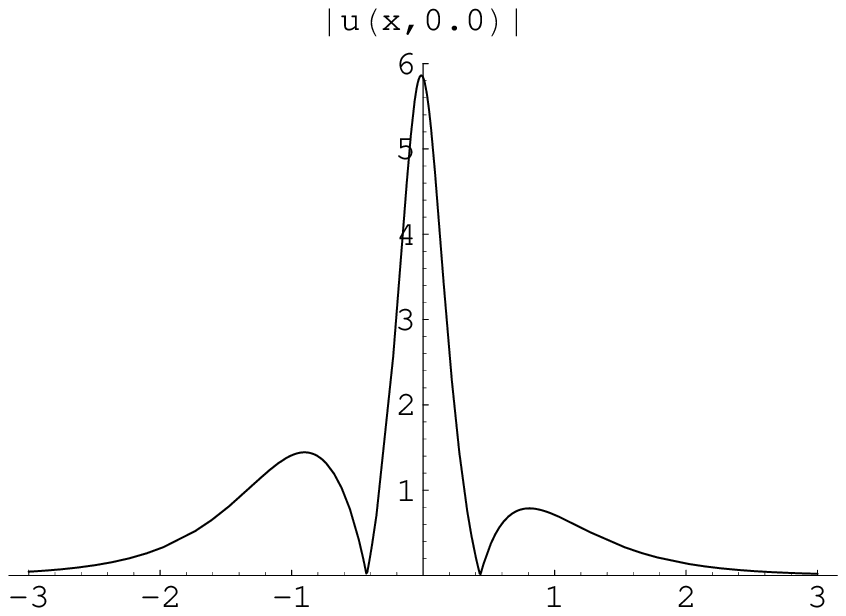,width=7 truein,height=4 truein}

\newpage

\psfig{figure=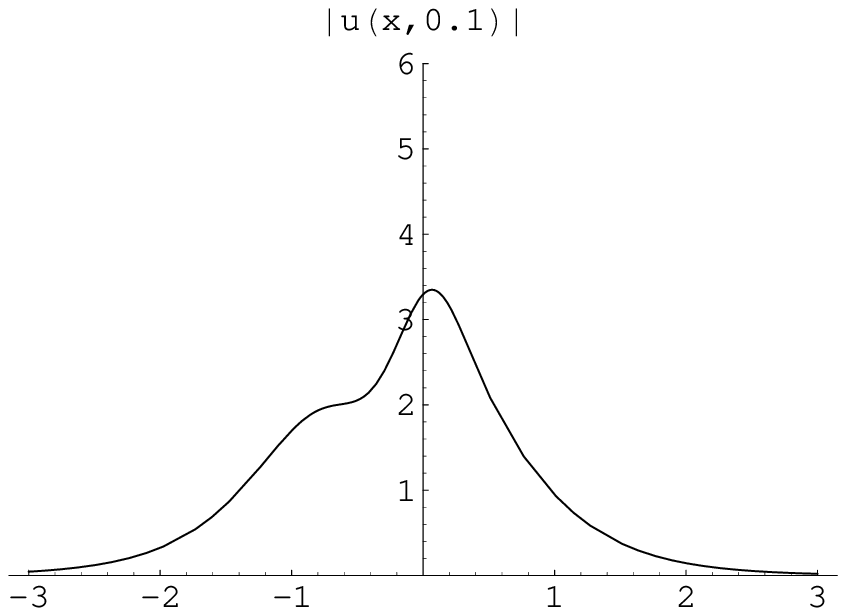,width=7 truein,height=4 truein}

\newpage

\psfig{figure=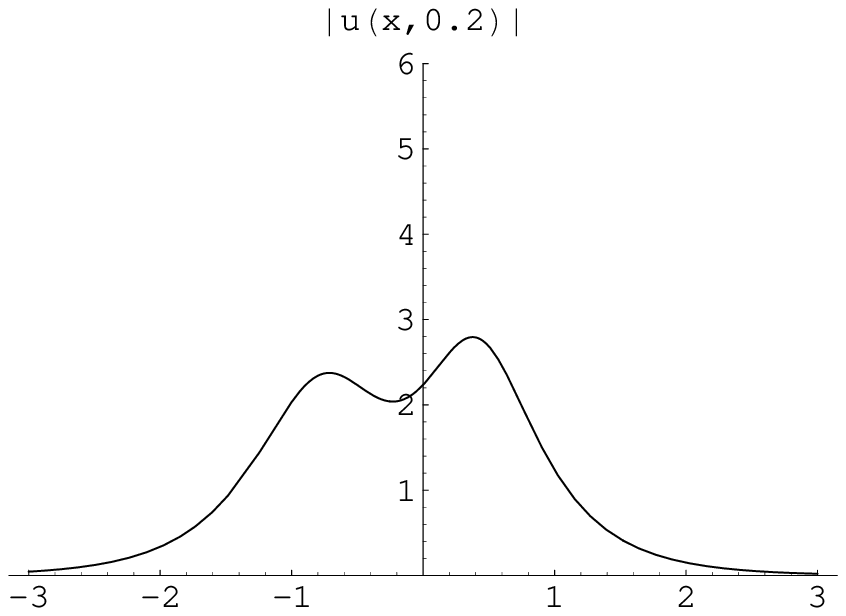,width=7 truein,height=4 truein}

\newpage

\psfig{figure=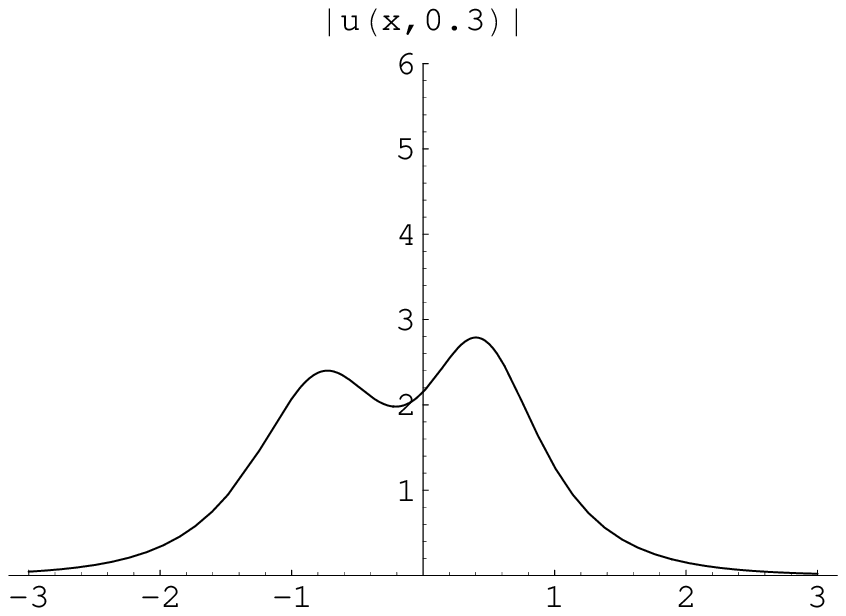,width=7 truein,height=4 truein}

\newpage
\psfig{figure=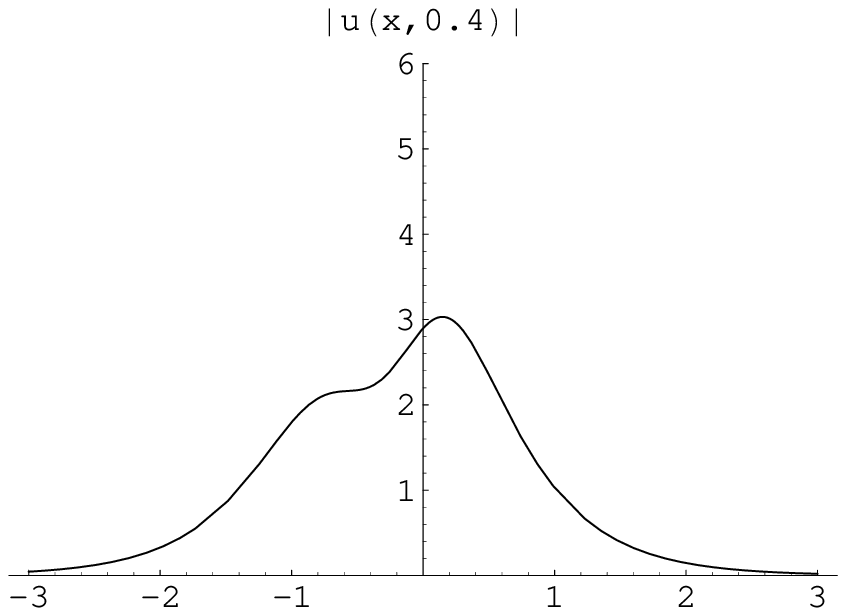,width=7 truein,height=4 truein}

\newpage
\psfig{figure=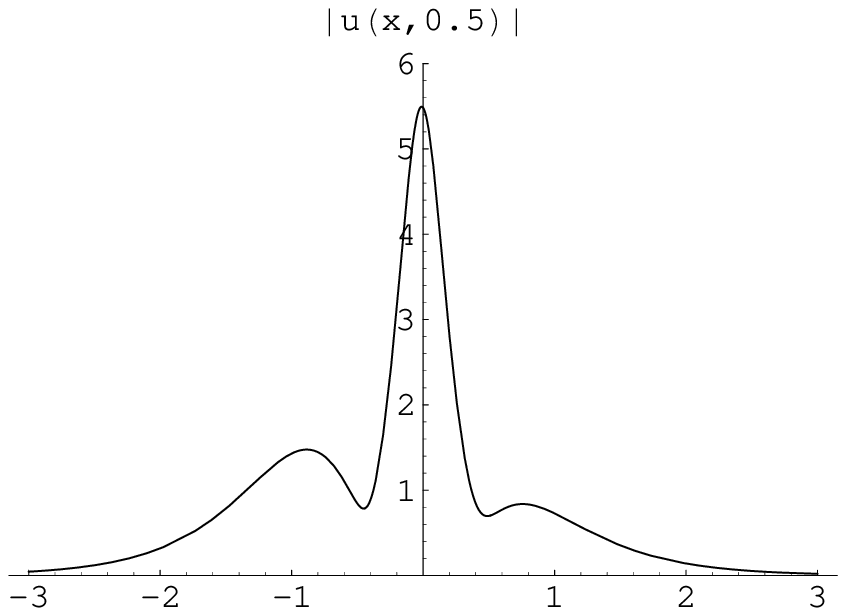,width=7 truein,height=4 truein}

\newpage

\centerline{{\bf Fig.~7.2} Snapshots of $|u(x,t)|$
of Example 7.3 at $t=-0.5,$ $-0.2,$
$-0.1,$ $0.0,$ $0.1,$ and $0.2.$}

\newpage
\psfig{figure=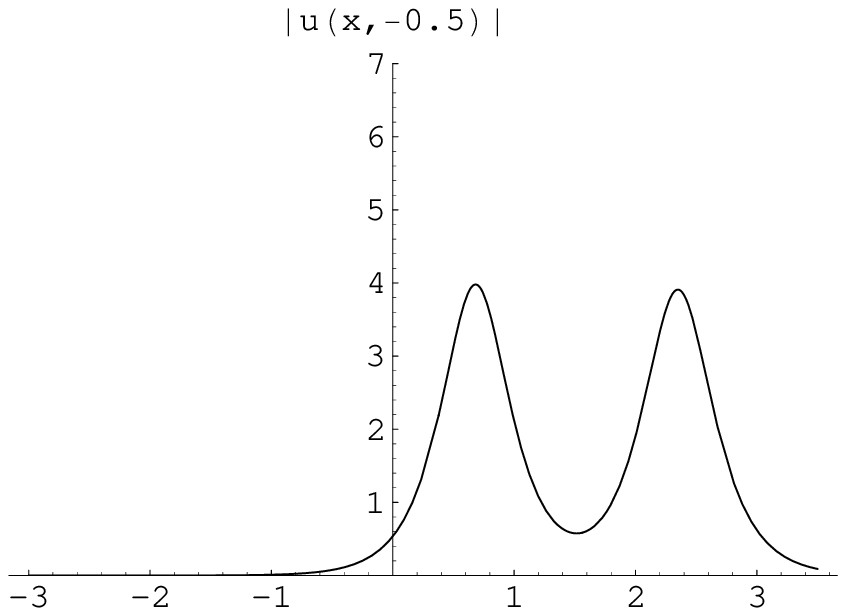,width=7 truein,height=4 truein}

\newpage
\psfig{figure=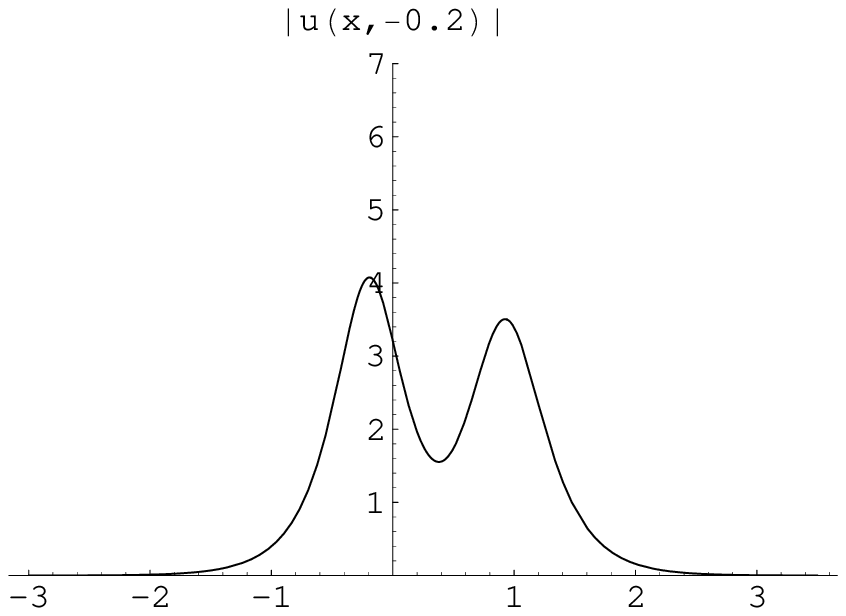,width=7 truein,height=4 truein}

\newpage
\psfig{figure=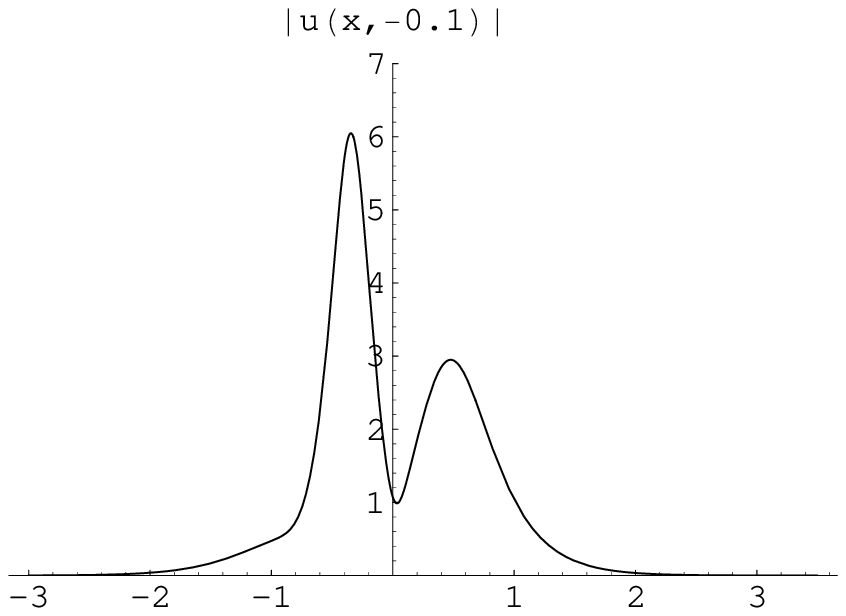,width=7 truein,height=4 truein}

\newpage
\psfig{figure=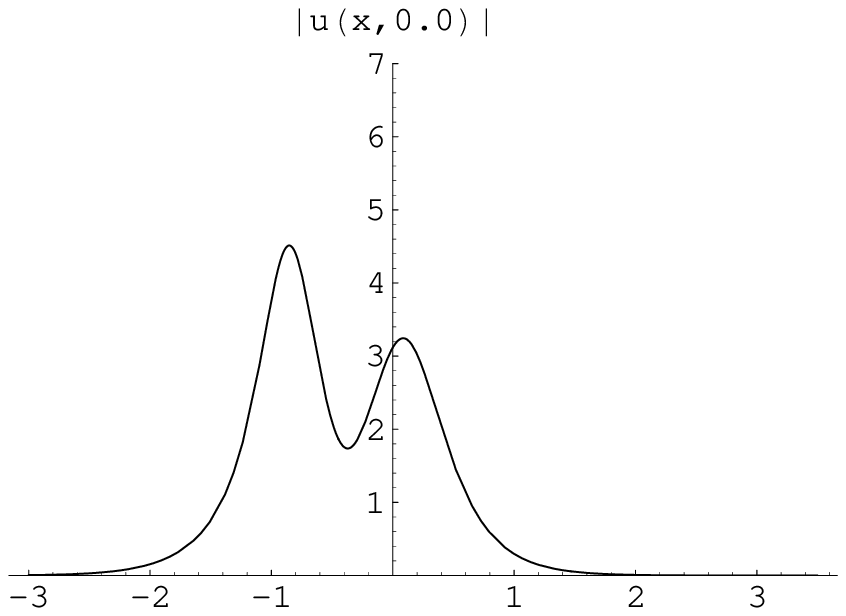,width=7 truein,height=4 truein}

\newpage
\psfig{figure=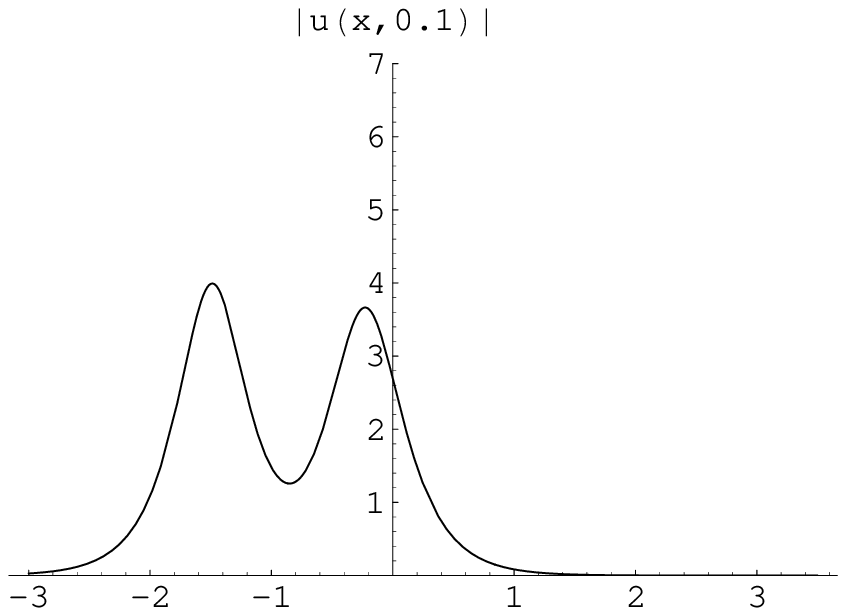,width=7 truein,height=4 truein}

\newpage
\psfig{figure=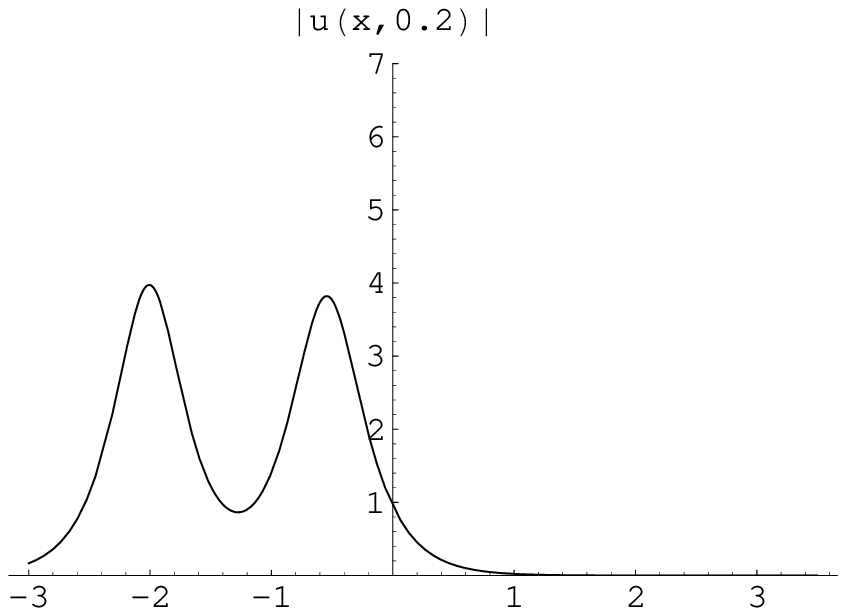,width=7 truein,height=4 truein}

\newpage

\centerline{{\bf Fig.~7.3} Snapshots of $|u(x,t)|$
of Example 7.4 at $t=0.0,$ $0.1,$
$0.2,$ $0.3,$ $0.4,$ and $0.5.$}

\newpage
\psfig{figure=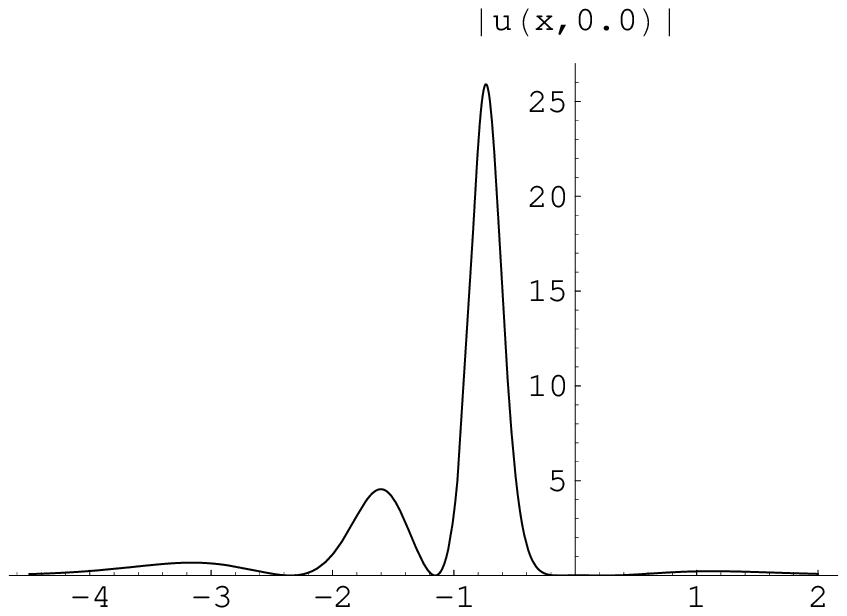,width=7 truein,height=4 truein}

\newpage
\psfig{figure=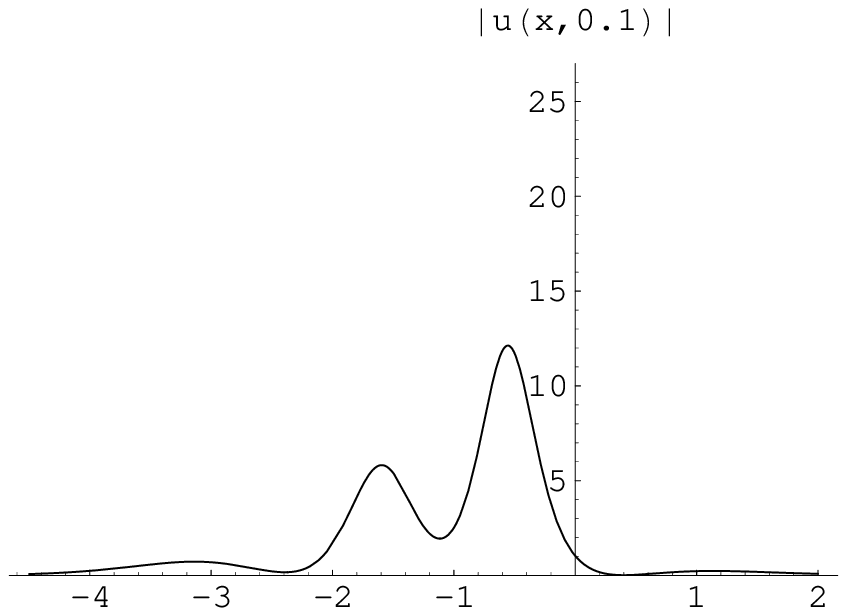,width=7 truein,height=4 truein}

\newpage
\psfig{figure=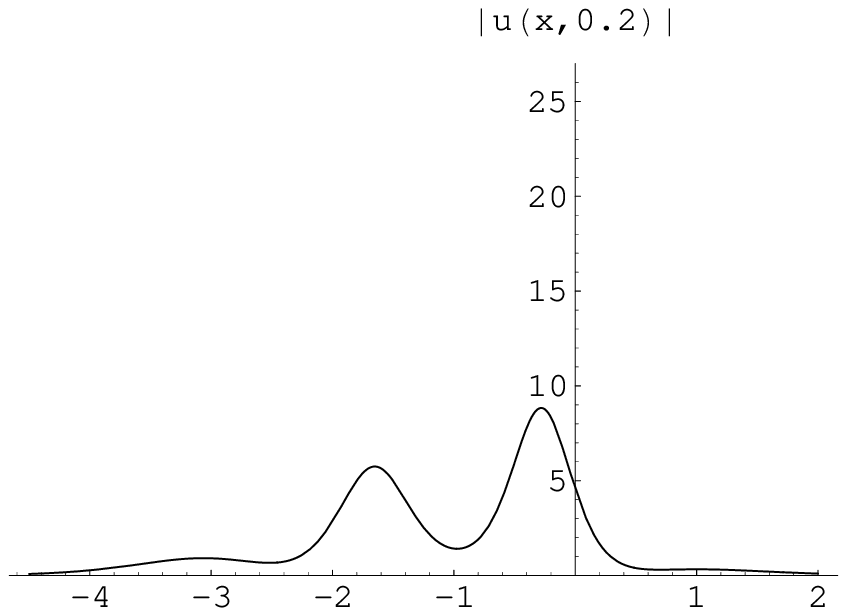,width=7 truein,height=4 truein}

\newpage
\psfig{figure=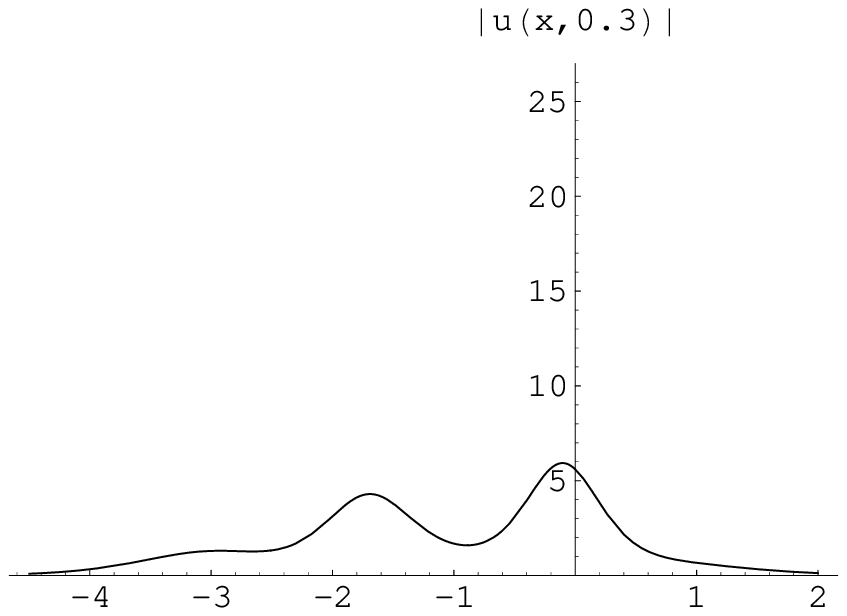,width=7 truein,height=4 truein}

\newpage
\psfig{figure=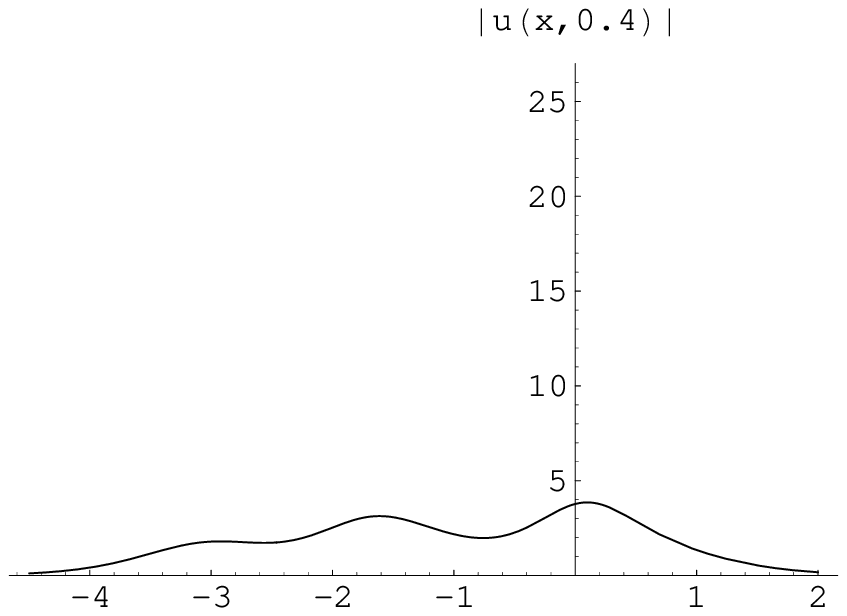,width=7 truein,height=4 truein}

\newpage
\psfig{figure=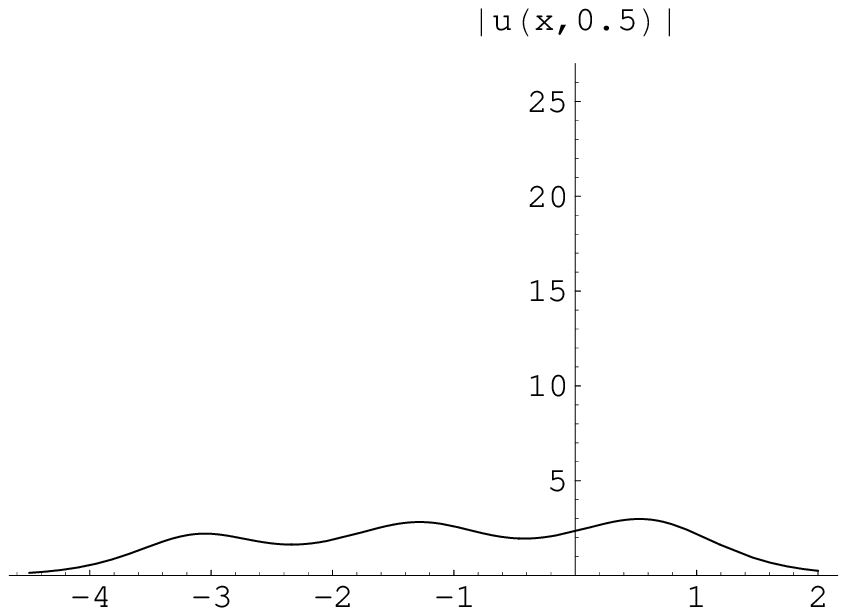,width=7 truein,height=4 truein}

\end